\documentclass[10pt,a4paper,twocolumn,aps,pra,showpacs,superscriptaddress,floatfix]{revtex4-1}
\usepackage[latin9]{inputenc}
\usepackage{amsmath}
\usepackage{amssymb}
\usepackage{graphicx}
\usepackage%[unicode=true,
{hyperref}
\usepackage{breakurl}

\makeatletter

\usepackage{dcolumn}% Align table columns on decimal point
\usepackage{bm}% bold math
\usepackage[pdftex]{color}

\usepackage{amsfonts}
\usepackage{amsthm}

\usepackage{rotate}
\let\oldmarginpar\marginpar
\renewcommand{\marginpar}[1]{\-\oldmarginpar[\raggedleft\tiny #1]%
{\raggedright\tiny #1}}

\def\KeyWord#1{$\backslash$\IfColor{$\!\!$\textRed{#1}\textBlack}{#1}$\!\!$}

\newcommand{\be}{\begin{equation} }
\newcommand{\ee}{\end{equation} }
\newcommand{\ba}{\begin{eqnarray} }
\newcommand{\ea}{\end{eqnarray} }
\newcommand{\n}{\nonumber \\ }

\newcommand{\mac}{\mathcal}

\newcommand{\bit}{\begin{itemize}}
\newcommand{\eit}{\end{itemize}}

\newcommand{\ns}{NbSe$_2$ }

\DeclareMathOperator{\sech}{sech}
\DeclareMathOperator{\sgn}{sgn}

\graphicspath{{Figures/}}

\makeatother

\begin{document}

\title{Crystalline nodal topological superconductivity and Bogolyubov Fermi surfaces in monolayer NbSe$_{2}$ }

\author{Daniel Shaffer}

\affiliation{School of Physics and Astronomy, University of Minnesota, Minneapolis,
Minnesota 55455, USA}

\author{Jian Kang}

\affiliation{National High Magnetic Field Laboratory, Florida State University,
Tallahassee, Florida 32310, USA}

\author{F. J. Burnell}

\affiliation{School of Physics and Astronomy, University of Minnesota, Minneapolis,
Minnesota 55455, USA}

\author{Rafael M. Fernandes}

\affiliation{School of Physics and Astronomy, University of Minnesota, Minneapolis,
Minnesota 55455, USA}

\date{\today}
\begin{abstract}
We present a microscopic calculation of the phase diagram of the Ising
superconductor NbSe$_{2}$ in presence of both in-plane magnetic
field and Rashba spin-orbit coupling (SOC). Repulsive interactions lead
to two distinct instabilities, in singlet- and triplet- interaction
channels. While we recover the previously predicted nodal topological superconducting state
 in the absence of Rashba SOC at large magnetic field with six pairs of nodes along \(\Gamma\)-\(M\) lines, 
 a finite Rashba SOC breaks the symmetry that protects these nodes and therefore generally lifts them,
 resulting in a topologically trivial phase. There is an exception
% In the regime of large fields, the topological character of the superconducting state depends strongly on the magnetic field direction.
when the field is applied along one of the three $\Gamma$-$K$
lines, however. In that case, a single mirror symmetry remains that can protect two pairs of nodes out of the original six,
resulting in a \emph{crystalline} topological superconducting phase. % is stabilized,
%whereas for other field directions the pairing state is topologically trivial.
Depending on the Cooper pairs' center-of-mass momentum, this superconducting state displays either Bogolyubov Fermi surfaces or point nodes. Moreover, a chiral topological superconducting phase with Chern number of 6 is realized in the regime of large Rashba SOC and dominant triplet interactions, spontaneously breaking time-reversal symmetry. 
\end{abstract}

\maketitle

\section{Introduction}

The observation of superconductivity (SC) in
1H monolayer transition metal dichalcogenides such as NbSe$_{2}$
and MoS$_{2}$ opens a new avenue to explore superconductivity in
systems with strongly coupled spin-orbital degrees of freedom \cite{Cao2015,Exp1MoS2,Exp2MoS2,MakPRL16,Exp3MoS2,Exp4MoS2,MakNat16,Exp1NbSe2,Exp1TMD,Hunt,Exp1TaS2}.
In contrast to their bulk counterparts, inversion symmetry
is broken in these monolayers, giving rise to an \textit{Ising} spin-orbit
coupling (SOC) that forces the spins to point out-of-plane \cite{MakNat16,Exp1NbSe2,LawPRB16,LawMay16II,LawMakNatMat18}.
This Ising SOC is believed to be responsible for the experimental
observation that the superconducting state survives up to remarkably
large in-plane magnetic fields, far beyond the usual Pauli limit \cite{Exp3MoS2,Exp4MoS2,MakNat16,Hunt,Houzet17,SigristPRL04}.

The combination of large Ising SOC, which lifts the spin degeneracy,
with multiple Fermi pockets has inspired considerable interest
in the potential for unconventional superconductivity in these materials
\cite{LawMay16,LawPRB16,Houzet17,AjiPRB17,YanasePRB17,Khodas18,Khodas19,Agterberg18,Agterberg17,LawPRL14,Oiwa18,YanasePRB17,KimNatCom17,Sticlet18}.
In gated MoS$_{2}$, which has four spin-split Fermi pockets centered
at the $\pm K$ points of the hexagonal Brillouin zone, repulsive
inter-band interactions can stabilize a fully gapped triplet SC state
\cite{LawPRL14,Oiwa18,YanasePRB17}. Chiral topological superconductivity
\cite{ReadGreen} both with and without large Rashba SOC has also
been predicted in MoS$_{2}$ \cite{LawPRL14,KimNatCom17,Oiwa19},
as has finite-momentum Cooper pairing \cite{KimNatCom17,pipStarykh}. In NbSe$_{2}$
and its close relative TaS$_{2}$, which have Fermi pockets centered
at the $\pm K$ and $\Gamma$ points, it was argued that for in-plane
magnetic fields larger than the Pauli limiting field, a nodal topological
SC state is realized, protected by an anti-unitary time-reversal like
symmetry and characterized by Majorana flat bands at the sample's
edges \cite{LawMay16,Agterberg18}.

\begin{figure}
\centering \includegraphics[width=0.99\columnwidth]{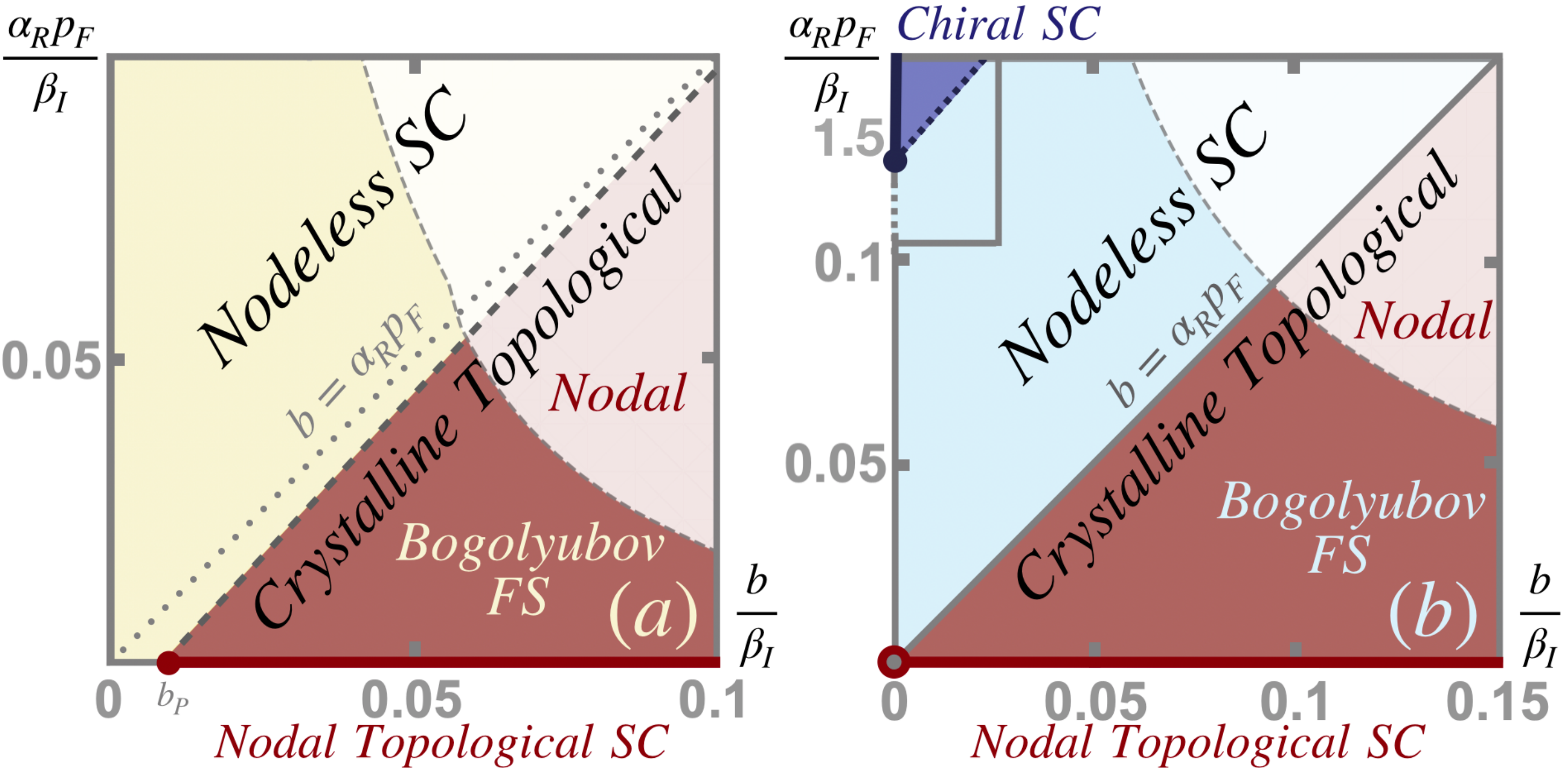}
\caption{Phase diagram for NbSe$_{2}$ as a function of the Rashba SOC $(\alpha_{R})$
and in-plane magnetic field $\mathbf{B}$ oriented along the $\Gamma$-$K$
direction, in units of the Ising SOC $\beta_{I}$. The leading SC instability at $\alpha_{R}=B=0$ is a singlet
extended $s$-wave state or triplet $f$-wave state in panel (a) and (b) respectively. Solid (dashed) lines indicate exact (approximate) phase
boundaries. Uniform SC becomes unstable in the light-shaded regions, but finite-momentum pairing remains possible. We discuss
the phases in more detail in Sec. \ref{SecIVB}.  We use parameter values given in Eqs. \ref{ParamFS} and \ref{ParamG}.}
\label{PhaseFig} 
\end{figure}

Despite the flurry of activity on this front, important questions
about the microscopic mechanism of unconventional SC and its stability in realistic experimental conditions remain unaddressed. Here, we go beyond phenomenological models and present a microscopic theory of superconductivity
in NbSe$_{2}$ that considers the most relevant repulsive electronic interactions involving low-energy fermions. Moreover, we include the simultaneous effects of an in-plane magnetic field $\mathbf{B}$ and Rashba SOC, with
energy scale $\alpha_{R}p_{F}$ ($p_{F}$ is the Fermi momentum). The latter is commonly present experimentally and can in principle be controlled by gating or by the choice of substrate. Importantly, it qualitatively changes the phase diagram: the nodal SC topological phase present at large fields \cite{LawMay16} is generally destroyed
by even a small Rashba SOC, as it lifts the nodes and breaks the time-reversal-like
symmetry protecting them. The exception is when
the $\mathbf{B}$ field is parallel to one of the $\Gamma$-$K$ directions: in that case, nodes located along the direction perpendicular to $\mathbf{B}$ are protected by mirror symmetry, resulting in a \emph{crystalline} topological SC phase that can be either nodal or exhibit protected Bogolyubov Fermi surfaces, depending on the momentum of the Cooper pairs.

Our analysis reveals two distinct $\left(B,\,\alpha_{R}\right)$ phase
diagrams, shown in Fig. \ref{PhaseFig}. For repulsive interactions, if the inter-band repulsion
coupling the $\Gamma$ and $\pm K$ Fermi pockets dominates, the SC
state for $B=\alpha_{R}=0$ is predominantly a singlet extended $s$-wave state
with nearly isotropic gaps of opposite signs at $\Gamma$ and $\pm K$
(Fig. \ref{PhaseFig}a). If the inter-band processes coupling the
$K$ and $-K$ Fermi pockets dominates, the dominant SC instability for $B=\alpha_{R}=0$
is towards a triplet $f$-wave state, characterized by isotropic gaps
of opposite signs at $K$ and $-K$, and a nodal gap at $\Gamma$. While the crystalline topological SC phase is present in
both phase diagrams for large enough fields, a distinct chiral topological $p\pm ip$ SC that
spontaneously breaks time-reversal symmetry is present for large $\alpha_{R}$ and $B\simeq0$ in the phase diagram of Fig. \ref{PhaseFig}(b).

The rest of the paper is organized as follows. In Sec. \ref{II} we introduce our microscopic model,
that includes Ising and Rashba SOC,
in-plane magnetic field, and all symmetry allowed spin-conserving interactions.
We analyze these interactions using a renormalization group (RG) analysis in Sec. \ref{III}, and find  
%At the high energy scale in RG, SOC and the magnetic field are negligible and we find
that (at energy scales where SOC and the magnetic field can be neglected) superconductivity is the only instability.
As a result, the RG at lower energy scales becomes equivalent to
a self-consistent mean field analysis.   In Sec. \ref{IV} we use this approach to study the superconducting phase diagram 
in the presence of SOC and magnetic field, and identify a new crystalline gapless topological superconducting phase at large in-plane magnetic field. 
%Assuming no pairing
%between electrons from inner and outer pairs, which is valid close to the phase transition,
%we project the interactions onto the split Fermi surfaces.
%Solving the resulting gap equation we find in Sec. \ref{SecIVA}, we go over the corresponding
%phase diagrams in Fig. \ref{PhaseFig} in Sec. \ref{SecIVB}. As
Because the Fermi surfaces are no longer symmetric under momentum inversion \(\mathbf{p}\rightarrow-\mathbf{p}\) in
the presence of both magnetic field and Rashba SOC, %the resulting superconducting phases become unstable in the opaque regions in Fig. \ref{PhaseFig}.
we also calculate the boundary of the region where uniform superconductivity becomes unstable in Sec. \ref{SecIVC}. 
 We show in Sec. \ref{V} that the gapless phase identified in Sec. \ref{IV} is a crystalline gapless topological phase, and analyze the
resulting boundary modes using a simple tight binding model.  We conclude in Sec. \ref{VI} by discussing possible experimental signatures of the
possible topological superconducting phases.  
A detailed discussion of the chiral state, as well as other technical supporting material, can be found in the appendices.  

\section{Microscopic model}\label{II}

\begin{figure}
\centering \includegraphics[width=0.9\columnwidth]{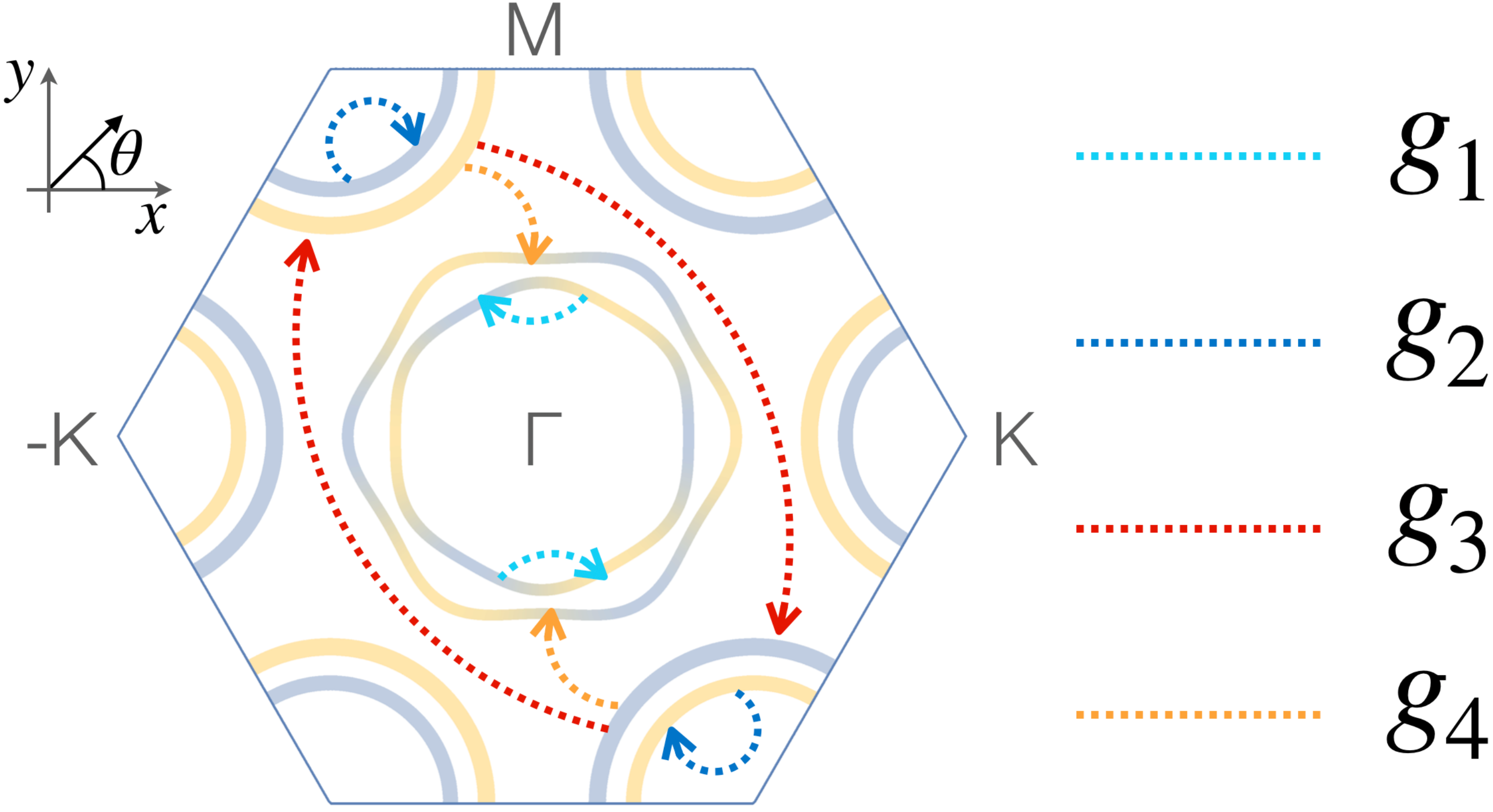} \caption{The Fermi surface of NbSe$_{2}$ in the presence of Ising SOC and
a weak Rashba SOC. The colors indicate the out-of-plane spin polarization
of each pocket. The arrows denote the four distinct types of repulsive
electronic interactions that contribute to the pairing instability. We used parameter values in Eq. (\ref{ParamFS}) for the non-interacting Hamiltonian, with \(b=0.5\beta_I\), and \(\alpha_R=0\); the Fermi surfaces are independent of the direction of the magnetic field in this case (\(b=0\), and \(\alpha_Rp_F=0.5\beta_I\) give the same Fermi surface). We picked values somewhat larger than reported in experiment to produce visible splitting at the \(\Gamma\) pocket. }
\label{FSFig} % \end{center}
 \end{figure}

The Fermi surface of undoped \ns is shown in Fig. \ref{FSFig}. The non-interacting Hamiltonian is given by: 
\begin{equation}\label{Eq:H0}
H_{0}=\sum_{\eta\mathbf{p}}\psi_{\eta,\mathbf{p}}^{\dag}\left[\epsilon_{\eta}({\bf p})+\beta_{\eta}({\bf p})\sigma^{z}+\alpha_{R}(\boldsymbol{\sigma}\times\mathbf{p})_{z}\right]\psi_{\eta,\mathbf{p}}
\end{equation}
where $\psi_{\eta,\mathbf{p}}^{\dag}=(d_{\eta,\mathbf{p}\uparrow}^{\dag},d_{\eta,\mathbf{p}\downarrow}^{\dag})$
and $d_{\eta,\mathbf{p}\alpha}^{\dagger}$ creates an electron at the pocket
$\eta$ with spin $\alpha=\uparrow,\,\downarrow$ and momentum $\mathbf{p}$
measured relative to the center of the pocket. 
The Fermi surface has
three pairs of spin-split hole pockets centered at the $\pm K,\Gamma$
points in the Brillouin zone, which we label by $\eta = \pm K, \Gamma$, using the convention that $-\Gamma\equiv\Gamma$. Here $\epsilon_{\eta}({\bf p})=-\frac{p^{2}}{2m_{\eta}}-\mu$
is the band dispersion, with $m_{K}=m_{-K}$.
The Ising SOC has the form $\beta_{\pm K}=\pm\beta_{I}$ near the
$\pm K$ points and $\beta_{\Gamma}=2\lambda p^{3}\cos3\theta$ near
the $\Gamma$ point, where $\theta$ is the angle measured relative
to the $\Gamma$-$K$ direction (see Fig. \ref{FSFig}). Although
Ising SOC vanishes along the $\Gamma$-$M$ lines, $\alpha_{R}$ does
not, so the spin-degeneracy is fully lifted on all Fermi pockets.
An in-plane magnetic field $\mathbf{B}$ adds a term $H_{\text{Zeeman}}=-\sum_{\eta\mathbf{p}}\psi_{\eta,\mathbf{p}}^{\dag}\left({\bf b}\cdot{\boldsymbol{\sigma}}\right)\psi_{\eta,\mathbf{p}}$,
where $\mathbf{b}\equiv \frac{1}{2}g_{L}\mu_{B}\mathbf{B}$ and $g_{L}$ is the
Land\'{e} g-factor; this also lifts the spin-degeneracy along the $\Gamma$-$M$ lines.
We then have
\begin{equation}\label{Eq:H0z}
H_{0}+H_{\text{Zeeman}}=\sum_{\eta\mathbf{p}}\psi_{\eta,\mathbf{p}}^{\dag}\left[\epsilon_{\eta}({\bf p})+\boldsymbol{\delta}_{\eta}({\bf p})\cdot\boldsymbol{\sigma}\right]\psi_{\eta,\mathbf{p}}
\end{equation}
where
\begin{equation}\label{deltaVec}
\boldsymbol{\delta}_{\eta}({\bf p})=\beta_\eta(\mathbf{p})\hat{\mathbf{z}}+\alpha_R(p_y\hat{\mathbf{x}}-p_x\hat{\mathbf{y}})+\mathbf{b}
\end{equation}
is the effective magnetic field seen by an electron with momentum \(\mathbf{p}\). The eigenvalues of the non-interacting Hamiltonian are therefore given by
\begin{equation} \label{Eq:Xidef}
\xi_{\eta\tau}(\mathbf{p})=\epsilon_\eta(\mathbf{p})+\tau\delta_\eta(\mathbf{p}) \ \ .
\end{equation}
where \(\delta_\eta=|\boldsymbol{\delta}_\eta|\), and the eigenstates are related to the electron annihilation operators $d_{\eta, {\bf p} \alpha}$ by the unitary 
transformation 
\begin{equation} 
c_{\eta,\mathbf{p}\tau}=U_{\eta\tau}^{\alpha}(\mathbf{p})d_{\eta,\mathbf{p}\alpha}
\label{Cbasis}
\end{equation}
Here $\tau=+1$ ($-1)$ on the outer (inner) spin-split Fermi surface,
$\alpha=1$ ($-1)$ for spin up (spin down) (see Fig. \ref{FSFig}), and we have defined 
\begin{eqnarray}\label{SOCbasis}
U_{\Gamma\tau}^{\alpha}(\mathbf{p}) & = & \sqrt{\frac{\delta_{\Gamma}+\tau\alpha(2\lambda p^{3}\cos3\theta_{{\bf p}})}{2\delta_{\Gamma}}}\left(\tau e^{-i\phi}\right)^{\frac{1+\alpha}{2}}\nonumber \\
U_{\pm K\tau}^{\alpha}(\mathbf{p}) & = & \sqrt{\frac{\delta_{K}\pm\tau\alpha\beta_{I}}{2\delta_{K}}}\left(\tau e^{-i\phi}\right)^{\frac{1+\alpha}{2}}
\end{eqnarray}
%{\color{blue}%[suggested alternative]:
%\begin{eqnarray}\label{SOCbasis}
%U_{\eta\tau}^{\alpha}(\mathbf{p}) & = & \sqrt{\frac{\delta_{\eta}+\tau\alpha\beta_{\eta}}{2\delta_{\eta}}}\left(\tau e^{-i\phi}\right)^{\frac{1+\alpha}{2}}
%\end{eqnarray}}
where
%{\color{blue}%[suggested alternative:]:
%\begin{eqnarray}\label{delta}
%\delta_{\eta}=|\boldsymbol{\delta}_\eta| & = & \sqrt{\beta_{\eta}^{2}+(\alpha_{R}p_{y}+b_{x})^{2}+(\alpha_{R}p_{x}-b_{y})^{2}}\nonumber 
%\end{eqnarray}
%}
\begin{eqnarray}\label{delta}
\delta_{\Gamma} & = & \sqrt{(2\lambda p^{3}\cos(3\theta))^{2}+(\alpha_{R}p_{y}+b_{x})^{2}+(\alpha_{R}p_{x}-b_{y})^{2}}\nonumber \\
\delta_{K} & = & \sqrt{\beta_{I}^{2}+(\alpha_{R}p_{y}+b_{x})^{2}+(\alpha_{R}p_{x}-b_{y})^{2}}\nonumber \\
e^{i\phi} & = & \frac{\alpha_{R}p_{y}+b_{x}+i(-\alpha_{R}p_{x}+b_{y})}{\sqrt{(\alpha_{R}p_{y}+b_{x})^{2}+(\alpha_{R}p_{x}-b_{y})^{2}}}
\end{eqnarray}
%where $\mathbf{p}$ is the momentum at the Fermi surface centered
%at $\Gamma$ or $\pm K$,
% $\mathbf{b}\equiv g_{L}\mu_{B}\mathbf{B}$,
%$\alpha_{R}$ is the Rashba SOC parameter, $\beta_{I}$ is the Ising
%SOC parameter at $K$, and $\lambda$ is the Ising SOC parameter at
%$\Gamma$. 

Experimentally,  the Ising SOC at \(K\) points $\beta_I$ is found to be much smaller than the bandwidth. 
Thus in the weak coupling theory used here, the overall energy scale is set by \(\beta_I\), and the momentum scale is set by the Fermi momentum $p_F$.  
To produce the plots in this paper, we took equal masses and chemical potentials at \(\Gamma\) and \(K\) points, and chose
\begin{equation}\label{ParamFS}
m=1.5 \frac{p_F^2}{\beta_I},\qquad \mu=-5\beta_I,\qquad \lambda=0.5\beta_I
\end{equation}
with $p$ shown in units of $p_F$.   This choice roughly matches the Fermi surfaces reported in \cite{LawMay16,MakNat16,Hunt,Khodas18}. %The momentum scale is set by  \(p_F=\sqrt{2m|\mu|}\), which is 
Experimentally,  for bulk NbSe\(_2\), \(p_F/\hbar = 0.5\) \AA\(^{-1}\) \cite{Borisenko, 1HCDW}. Note that the estimated value of $\beta_I$ is of the order of $40$ meV \cite{MakNat16,Hunt}. Thus, the maximum values of $b$ in our phase diagrams of Fig. \ref{PhaseFig} correspond roughly to fields of $70$ T and $100$ T. Although the largest values in this range are most likely above the experimental critical in-plane field value, given the uncertainty in the microscopic parameters, we opt to extend the phase diagrams across a wide range of magnetic field values.
 %For a sense of scale of the rest of the parameters, Hunt et al. \cite{Hunt} experimentally find \(\beta_I\approx40\) meV for NbSe\(_2\) (and \(\approx100\) meV for the chemically similar TaS\(_2\)), and report a critical field \(b\approx3\) meV (corresponding to about 30 T), to be compared to the band width of \(\approx1\) eV \cite{MakNat16,Hunt,Khodas18}. They also find \(\alpha_R p_F\approx 6\) meV based on a fit of their data to a phenomenological model first used for MoS\(_2\) in \cite{Exp3MoS2} and \cite{Exp4MoS2}. Estimates of \(\lambda p_F^3\) are not available, but we expect this parameter to be somewhat smaller than \(\beta_I\). 

\subsection*{Interactions Near the Fermi Surfaces}

Symmetry constrains the possible spin-conserving, momentum-direction-independent electronic interactions between the low-energy fermionic operators to eight.  
Of these, the four interactions that contribute to superconductivity are (see Fig. \ref{FSFig}): intra-pocket density-density interactions
involving the $\Gamma$ ($g_{1}$) and the $\pm K$ ($g_{2}$) pockets;
and inter-pocket pair-hopping interactions between $K$ and $-K$
($g_{3}$) and between $\Gamma$ and $\pm K$ ($g_{4}$). %check naming convention
Thus we consider the following interacting Hamiltonian:
%(momentum indices are suppressed for simplicity):
%\begin{align}
%H_{\text{Int}}=\frac{g_{1}}{2}d_{\Gamma s}^{\dagger}d_{\Gamma s'}^{\dagger}d_{\Gamma s'}d_{\Gamma s}+\frac{g_{2}}{2}d_{Ks}^{\dagger}d_{-Ks'}^{\dagger}d_{-Ks'}d_{Ks}+\nonumber \\
%+\frac{g_{3}}{2}d_{Ks}^{\dagger}d_{-Ks'}^{\dagger}d_{Ks'}d_{-Ks}+\frac{g_{4}}{2}d_{\pm Ks}^{\dagger}d_{\mp Ks'}^{\dagger}d_{\Gamma s'}d_{\Gamma s}+\mathrm{h.c.}\label{H_int}
%\end{align}
\begin{align} 
H_{\text{Int}}= & V_{\Gamma;\Gamma}^{\alpha'\beta';\alpha\beta}(\mathbf{p};\mathbf{k})\ d_{\Gamma,\mathbf{p}\alpha}^{\dagger}d_{\Gamma,-\mathbf{p}\beta'}^{\dagger}d_{\Gamma,\mathbf{k}\alpha'}d_{\Gamma,-\mathbf{k}\beta'}+\nonumber \\
 & V_{\pm K;\pm K}^{\alpha'\beta';\alpha\beta}(\mathbf{p};\mathbf{k})\ d_{\pm K,\mathbf{p}\alpha}^{\dagger}d_{\mp K,\mathbf{-p}\beta}^{\dagger}d_{\pm K,\mathbf{k}\alpha'}d_{\mp K,\mathbf{-k}\beta'}+\nonumber \\
 & V_{\pm K;\mp K}^{\alpha'\beta';\alpha\beta}(\mathbf{p};\mathbf{k})\ d_{\pm K,\mathbf{p}\alpha}^{\dagger}d_{\mp K,\mathbf{-p}\beta}^{\dagger}d_{\mp K,\mathbf{k}\alpha'}d_{\pm K,\mathbf{-k}\beta'}+\nonumber \\
 & V_{\Gamma;\pm K}^{\alpha'\beta';\alpha\beta}(\mathbf{p};\mathbf{k})\ d_{\pm K,\mathbf{p}\alpha}^{\dagger}d_{\mp K,\mathbf{-p}\beta}^{\dagger}d_{\Gamma,\mathbf{k}\alpha'}d_{\Gamma,\mathbf{-k}\beta'}+\mathrm{h.c.}\label{H_int}
\end{align}
Accounting for the anti-symmetric nature of the fermion
operators (and including all Hermitian conjugates), the uniform part
of the interactions can be separated into singlet and triplet interaction
channels, as follows:

\begin{eqnarray} \label{Eq:Ints_SpinChannel}
\left[V^{s}\right]_{\Gamma;\Gamma}^{\alpha'\beta';\alpha\beta} & = & \frac{g_{1}}{2}(i\sigma^{y})^{\alpha\beta}(i\sigma^{y})^{\alpha'\beta'}  \\
\left[V^{s}\right]_{\Gamma;\pm K}^{\alpha'\beta';\alpha\beta} & = & \pm \frac{g_{4}}{2}(i\sigma^{y})^{\alpha\beta}(i\sigma^{y})^{\alpha'\beta'}\nonumber \\
\left[V^{s}\right]_{\pm K;\pm K}^{\alpha'\beta';\alpha\beta} & = & \frac{1}{4}(g_{2}+g_{3})(i\sigma^{y})^{\alpha\beta}(i\sigma^{y})^{\alpha'\beta'}\nonumber \\
\left[V^{t}\right]_{\pm K;\pm K}^{\alpha'\beta';\alpha\beta} & = & \frac{1}{4}(g_{2}-g_{3})\sum_{j=x,y,z}(\sigma^{j}i\sigma^{y})_{\alpha\beta}^{*}(\sigma^{j}i\sigma^{y})^{\alpha'\beta'}\nonumber 
\end{eqnarray}
Since $V_{K, K}$ and $V_{ K, -K}$ are related by interchanging the spin indices $\alpha', \beta'$, combined with an overall minus sign for interchanging two fermion operators, in this representation we have
\begin{eqnarray}
\left[V^{s}\right]_{\pm K;\mp K}^{\alpha'\beta';\alpha\beta} & = &\left[V^{s}\right]_{\pm K;\pm K}^{\alpha'\beta';\alpha\beta}  %=  \frac{1}{2}(g_{2}+g_{3})(i\sigma^{y})^{\alpha\beta}(i\sigma^{y})^{\alpha'\beta'}
 \nonumber \\
\left[V^{t}\right]_{\pm K;\mp K}^{\alpha'\beta';\alpha\beta} & = & -\left[V^{t}\right]_{\pm K;\pm K}^{\alpha'\beta';\alpha\beta} %= - \frac{1}{2}(g_{2}-g_{3})\sum_{i=x,y,z}(\sigma^{i}i\sigma^{y})_{\alpha\beta}^{*}(\sigma^{i}i\sigma^{y})^{\alpha'\beta'}\nonumber 
\end{eqnarray}

From Eq. (\ref{Eq:Ints_SpinChannel}), we see that $V_{\pm K, \pm K}$ (and thus $V_{\pm K, \mp K}$)  have contributions in both the
singlet channel (labeled $s$) and the triplet channel (labeled $t$),
while for momentum-direction-independent interactions, $V_{\Gamma,\Gamma}$
and $V_{\Gamma,K}$ have contributions only in the singlet channel.
In addition to these interactions, in order to
ensure that the gap on the $\Gamma$ pocket does not artificially vanish in the
triplet regime, we also include weak (but symmetry-allowed) momentum-direction
dependent interactions:
\begin{eqnarray}\label{tripInt}
\left[V^{t}(\mathbf{p};\mathbf{k})\right]_{\Gamma;\Gamma}^{\alpha'\beta';\alpha\beta} & = & \frac{g_{1}^{t}}{2}\cos3\theta_{\mathbf{k}}\cos3\theta_{\mathbf{p}}(\boldsymbol{\sigma}i\sigma^{y})^*_{\alpha\beta}\cdot(\boldsymbol{\sigma}i\sigma^{y})^{\alpha'\beta'} \nonumber \\
\left[V^{t}(\mathbf{p};\mathbf{k})\right]_{\Gamma;\pm K}^{\alpha'\beta';\alpha\beta} & = & \pm\frac{g_{4}^{t}}{\sqrt{2}}\cos3\theta_{\mathbf{k}}(\boldsymbol{\sigma}i\sigma^{y})_{\alpha\beta}^{*}\cdot(\boldsymbol{\sigma}i\sigma^{y})^{\alpha'\beta'}\nonumber \\
\end{eqnarray}
where $\theta_{\mathbf{k}}$ refers to the angle of the momentum on
the $\Gamma$ pocket. Note that all higher harmonics can be included in a similar fashion, but they do not qualitatively affect our conclusions. We take $|g_{i}^{t}|\ll|g_{i}|$,
so these interactions have a negligible effect on whether the
system enters the singlet or triplet regime.

Note that there are four other possible interactions, which we do not include in our analysis here,
\begin{align}\label{HInt2}
H_{\text{Int}}=\frac{g_{5}}{2}d_{K \alpha}^{\dagger}d_{K \beta}^{\dagger}d_{K \beta}d_{K \alpha}+\frac{g_{6}}{2}d_{-K\alpha}^{\dagger}d_{\Gamma \beta}^{\dagger}d_{\Gamma \beta}d_{-K \alpha}+\nonumber \\
+\frac{g_{7}}{2}d_{-K\alpha}^{\dagger}d_{\Gamma \beta}^{\dagger}d_{-K \beta}d_{\Gamma \alpha}+\frac{g_{8}}{2}d_{-K\alpha}^{\dagger}d_{\Gamma \beta}^{\dagger}d_{K \beta}d_{K \alpha}+\mathrm{h.c.}
\end{align}
where we omitted momentum indices and symmetry related terms for simplicity. These interactions decouple from $g_1,...g_4$, and hence need not be included when analyzing the possible superconducting instabilities. They can in principle give rise to a pair density wave (PDW) which may compete with superconductivity, depending on the microscopic values of $g_5,...g_8$.  We defer analysis of this possibility to future work.

\section{Renormalization Group and Superconducting Instability}\label{III}

To determine which instabilities are favored by the interactions above, we perform a parquet RG analysis, keeping only the dominant momentum-direction-independent interactions of Eq. (\ref{Eq:Ints_SpinChannel}).  This approach is appropriate for the situation when the Fermi energy is small compared to the bandwidth, as one integrates out states from energies of the order of the bandwidth to energies of the order of the Fermi energy \cite{Chubukov16,Vafek12,Scheurer15,Mengxing18}; in the case of NbSe$_2$ the Fermi surfaces can be made small by controlling the gate voltage. Below the Fermi energy, the different channels decouple, and in the absence of nesting, the only logarithmic instability is the superconducting one. Because the energy scale of the Ising SOC is smaller than the Fermi energy, we perform the parquet RG in the absence of SOC or magnetic fields. These terms however are relevant as we move to energy scales below the Fermi energy, which we explore in Section \ref{IV}.  

\subsection{RG Flow Equations}

It is convenient to rescale the coupling constants %for the singlet {\color{cyan}\(s\) and triplet \(t\)} channels 
by the density of states (DOS) \(N_\eta=\frac{m_\eta}{2\pi}\) of the \(\eta\) pocket (by symmetry \(N_K=N_{-K}\)): 
\begin{eqnarray}\nonumber\label{gs}
&\tilde{g}_1^{(s)}=N_\Gamma g_1,\quad \tilde{g}_4^{(s)}=\sqrt{N_\Gamma N_K} g_4, \quad \tilde{g}_{23}^{(s)}=N_K \frac{g_2+g_3}{2}\\
&\tilde{g}_1^{(t)}=N_\Gamma g_1^{t}, \quad \tilde{g}_4^{(t)}=\sqrt{N_\Gamma N_K} g_4^{t}, \quad \tilde{g}_{23}^{(t)}=N_K \frac{g_2-g_3}{2}\nonumber\\
\end{eqnarray}
We use the standard parquet RG procedure \cite{Chubukov16,Vafek12,Scheurer15,Mengxing18}. Since all pockets are hole pockets, only ladder diagrams contribute to logarithmic instabilities at one-loop order.  In the absence of SOC the RG flow equations for for the singlet $(a=s)$ and triplet $(a=t)$ channels decouple:
\begin{eqnarray}\label{Eq:RG}
\dot{\tilde{g}}_1^{(a)}&=-\left(\tilde{g}^{(a)}_1\right)^2-2\left(\tilde{g}^{(a)}_4\right)^2\\
\dot{\tilde{g}}_{23}^{(a)}&=-2\left(\tilde{g}^{(a)}_{23}\right)^2-\left(\tilde{g}^{(a)}_4\right)^2\\
\dot{\tilde{g}}_4^{(a)}&=-(\tilde{g}^{(a)}_1+2\tilde{g}^{(a)}_{23})\tilde{g}_4^{(a)}
\end{eqnarray}
where the dot indicates a derivative with respect to the RG scale \(s\) determined by the pairing susceptibility via
\begin{equation} \label{Eq:Pi}
-N_\eta s=\Pi_\eta=T\sum_n \int_{d\Lambda} G^{(0)}_\eta(i\omega_n,\mathbf{Q})G^{(0)}_{-\eta}(-i\omega_n,-\mathbf{Q})\frac{d^2Q}{(2\pi)^3}
\end{equation}
Here \(\omega_n\) are Matsubara frequencies, and the momentum integral is restricted to a thin shell at energy \(\Lambda\) and of thickness \(d\Lambda\equiv\Lambda s\).  $G^{(0)}_\eta(i\omega_n,\mathbf{Q})=(i\omega_n-\epsilon_\eta(\mathbf{Q}))^{-1}$ is the bare normal state Green's function.
%\begin{equation}
%G^{(0)}_\eta(i\omega_n,\mathbf{Q})=\frac{1}{i\omega_n-\epsilon_\eta(\mathbf{Q})}
%\end{equation}
 
%In cylindrical co-ordinates 
Eq. (\ref{Eq:RG}) has an analytic solution, which can be obtained by switching to a cylindrical coordinate system in the \(\tilde{g}_1^{(a)}\), \(\tilde{g}_{23}^{(a)}\), and  \(\tilde{g}_4^{(a)}\) parameter space:
\begin{equation}
z^{(a)}=2\tilde{g}^{(a)}_{23}+\tilde{g}^{(a)}_1\nonumber
\end{equation}
\begin{equation}
r^{(a)}\cos\theta^{(a)}=\tilde{g}^{(a)}_4\qquad r^{(a)}\sin\theta^{(a)} = 2\tilde{g}^{(a)}_{23}-\tilde{g}^{(a)}_1 \nonumber
\end{equation}
Eq. (\ref{Eq:RG}) becomes:
\begin{align}
\dot{z}^{(a)}&=-\frac{\left(z^{(a)}\right)^2}{2}-\frac{\left(r^{(a)}\right)^2}{2}(1+7\cos^2\theta^{(a)})\\
\dot{r}^{(a)}&=-r^{(a)}z^{(a)}\\
\dot{\theta}^{(a)}&=0
\end{align}
We see that one channel, parametrized by \(\theta^{(a)}\), is not renormalized within one-loop. The other two channels decouple,  according to:
\begin{equation}
\dot{\gamma}^{(a\pm)}=%-\frac{1}{2}\frac{d}{ds}(z^{(a)}\mp r^{(a)}\sqrt{1+7\cos^2\theta^{(a)}})=
(\gamma^{(a\pm)})^2
\end{equation}
where, in terms of the coupling constants $\tilde{g}^a$, we have:
\begin{eqnarray} \label{Eq:gamma}
\gamma^{(a\pm)}&=&-\frac{1}{2}\left(z^{(a)}\mp r^{(a)}\sqrt{1+7\cos^2\theta^{(a)}}\right) \\
&=&- \frac{1}{2} \tilde{g}^{(a)}_1-\tilde{g}^{(a)}_{23}\pm \frac{1}{2} \sqrt{\left(\tilde{g}^{(a)}_1-2\tilde{g}^{(a)}_{23}\right)^2+8\left(\tilde{g}^{(a)}_4\right)^2}\nonumber
\end{eqnarray}
When $\gamma^{(a\pm)} >0$, the associated coupling constant flows to $\infty$.  Since $\gamma^{(a +)}>\gamma^{(a -)}$, the former always diverges faster than the latter, so we need only analyze the $\gamma^{(a+)}$ solution.  When \(N_\eta=N\) are all equal, these can be expressed:
\begin{equation}\label{gamma0}
\gamma^{(s)}=\gamma^{(s+)}=- \frac{N}{2} (g_1+g_2+g_3) +\frac{N}{2} \sqrt{(g_1-g_2-g_3)^2+8g^2_4}
\end{equation}
in the singlet channel and
\begin{equation}\label{gammat}
\gamma^{(t)}=\gamma^{(t+)}= \frac{N}{2} ( g_3 -g_2 + \left|g_2-g_{3}\right| ) 
\end{equation}
in the triplet channel. %{\color{blue}Thus, for repulsive interactions, a SC state is realized when the the inter-band repulsions $g_3$ and $g_4$ dominate over the intraband repulsions $g_1$ and $g_2$. In this case, the singlet instability dominates for large $g_4$, while the triplet instability dominates for large $g_3$.}%moved last sentence from below

 \subsection{Superconducting Instability}
 
Having determined the RG flow of the coupling constants, we now discuss which instabilities they cause. To do so, we introduce vertices associated with different types of electronic order
%. The superconducting vertices are associated with the gap function  
%{\color{blue}$\left[\Delta_{\eta}\left(\mathbf{p}\right)\right]_{\alpha\beta}\propto\left\langle d_{\eta,\mathbf{p}\alpha}d_{-\eta,-\mathbf{p}\beta}\right\rangle $ (\(\eta=\Gamma,\pm K\), with \(-\Gamma=\Gamma\) begin understood).}
 $\left(\Delta_{\Gamma}\left(\mathbf{p}\right)\right)_{\alpha\beta}\propto\left\langle d_{\Gamma,\mathbf{p}\alpha}d_{\Gamma,-\mathbf{p}\beta}\right\rangle $ at the $\Gamma$ pocket,
and $\left(\Delta_{\pm K}\left(\mathbf{p}\right)\right)_{\alpha\beta}\propto\left\langle d_{\pm K,\mathbf{p}\alpha}d_{\mp K,-\mathbf{p}\beta}\right\rangle $
%\varepsilon was no defined, and would imply PDW vertices
at the $\pm K$ pockets.   
Here $\alpha\, \beta$ are spin indices.
%$\varepsilon,\,\varepsilon'=\pm1$ are valley indices. 
In the absence of SOC the corresponding superconducting pairing vertices can be decomposed into singlet and triplet channels as follows:
\begin{eqnarray} \label{Deltas}
\Delta^{(s)}_\eta(\mathbf{p})&=& D^{(s)}_\eta i\sigma^y\nonumber\\
\Delta^{(t)}_\Gamma(\mathbf{p})&=&D^{(t)}_\Gamma \sqrt{2}\cos3\theta_{\Gamma,\mathbf{p}}\hat{\mathbf{d}}\cdot\boldsymbol{\sigma}\  i\sigma^y\nonumber\\
\Delta^{(t)}_{\pm K}(\mathbf{p})&=& \pm D^{(t)}_K\hat{\mathbf{d}}\cdot \boldsymbol{\sigma}\ i\sigma^y
\end{eqnarray}
where %\(j=x,y,z\) denotes the three different components of the triplet SC channel and
\(\theta_{\Gamma,\mathbf{p}}\) is the angle about the \(\Gamma\) Fermi surface, \(D^{(a)}_\eta\) are momentum-independent coefficients, and the (unit) \(\hat{\mathbf{d}}\) vector is the same on \(\Gamma\) and \(K\) pockets. Note that \(\Delta^{(a)}_\eta(\mathbf{p})=-\left[\Delta^{(a)}_{-\eta}(-\mathbf{p})\right]^T\) due to the anti-commutation relations of fermionic creation and annihilation operators. In particular, $\Delta_{K}$ and $\Delta_{-K}$ are not independent gap functions.
%; however, it is convenient to introduce a valley index (associated with the matrix $\tau^z$ above) to keep track of both, since the relationship is different in the singlet and triplet channels.    

The one-loop vertex flow correction is then
\begin{equation}\label{Eq:VertexFlow}
\left[\delta\Delta_\eta(\mathbf{p})\right]_{\alpha\beta}=\Pi_{\eta'}\int\left[V(\mathbf{k;p})\right]_{\eta;\eta'}^{\alpha\beta;\alpha'\beta'}\left[\Delta_{\eta'}(\mathbf{k})\right]_{\alpha'\beta'}\frac{d\theta_{\eta',\mathbf{k}}}{2\pi}
\end{equation}
with a sum over repeated indices implied. This reduces to a system of \(2\times2\) equations for the \(D^{(a)}_\eta\) coefficients:
\begin{equation} 
\label{Eq:Dgaps}
\frac{d}{ds}\left(\begin{array}{c}
D^{(a)}_\Gamma\\
D^{(a)}_{K}
\end{array}\right)=-\left(\begin{array}{ccc}
\tilde{g}^{(a)}_1 & 2\tilde{g}^{(a)}_{4}\sqrt{\frac{N_K}{N_\Gamma}}\\
\tilde{g}^{(a)}_4\sqrt{\frac{N_\Gamma}{N_K}} & 2\tilde{g}^{(a)}_{23}
\end{array}\right)\left(\begin{array}{c}
D^{(a)}_\Gamma\\
D^{(a)}_{K}
\end{array}\right)
\end{equation}
The eigenvalues of the matrix correspond precisely to the decoupled RG effective couplings \(\gamma^{(a\pm)}\); thus, $ \gamma^{(a +)}>0$ implies a SC transition in the corresponding $a$ channel. 

%Focusing on pairing with zero center-of-mass momentum
%forces $\varepsilon$ and $\varepsilon'$ to be opposite. 
Even when all $g_{j}$ interactions are purely repulsive, this leads to  two possible
SC instabilities, provided that one of the inter-pocket interactions,
$g_{3}$ or $g_{4}$, overcomes the intra-pocket repulsion promoted
by $g_{1}$ and $g_{2}$  -- see Eqs (\ref{gamma0}-\ref{gammat}).
%While Ising SOC allows mixture of singlet and triplet pairing, we will focus on the leading energetically-favored instabilities.
When $g_{4}$ is dominant, the resulting
SC state is a singlet $s$-wave, with isotropic gaps \(\Delta_\eta^{(s)}\).
%$\left(\Delta_{\Gamma}\left(\mathbf{p}\right)\right)_{\alpha\beta}=\Delta_{\Gamma,0}\left(i\sigma_{y}\right)_{\alpha\beta}$
%and $\left(\Delta_{K}\left(\mathbf{p}\right)\right)_{\alpha\beta}^{\varepsilon\varepsilon'}=\Delta_{K,0}\left(i\sigma^{y}\right)_{\alpha\beta}\left(\tau^{x}\right)_{\varepsilon\varepsilon'}$.
%Here, $\boldsymbol{\sigma}$ and $\boldsymbol{\tau}$ are Pauli matrices
%in spin and valley spaces, respectively.
% Because the two gaps have
%opposite signs, $\mathrm{sgn}\left(\Delta_{\Gamma,0}\right)=-\mathrm{sgn}\left(\Delta_{K,0}\right)$, this is
With repulsive interactions, $\mathrm{sgn}\left[D_{\Gamma}^{(s)}\right]=-\mathrm{sgn}\left[D_{K}^{(s)}\right]$,
the so-called extended $s$-wave or $s^{\pm}$-wave state,
previously proposed to be realized e.g. in iron pnictides \cite{Mazin11}
and strontium titanate \cite{Trevisan18}. 

In contrast, when $g_{3}$ is the dominant interaction, the SC instability
is towards a triplet $f$-wave state, characterized by %$\left[\Delta_{\eta}\left(\mathbf{p}\right)\right]_{\alpha\beta}=\left[\left(\mathbf{d}_{\eta}\left(\mathbf{p}\right)\cdot\boldsymbol{\sigma}\right)i\sigma^{y}\right]_{\alpha\beta}$.
$\left(\Delta_{\Gamma}\left(\mathbf{p}\right)\right)_{\alpha\beta}\propto \cos3\theta_{\Gamma,\mathbf{p}} \left[ \left(\hat{\mathbf{d}} \cdot\boldsymbol{\sigma}\right)i\sigma^{y}\right]_{\alpha\beta}$
and $\left(\Delta_{\pm K}\left(\mathbf{p}\right)\right)_{\alpha\beta} \propto \left[\left(\hat{\mathbf{d}} \cdot\boldsymbol{\sigma}\right)i\sigma^{y}\right]_{\alpha\beta} $.
Because spin is conserved in the absence of SOC and magnetic field, the $\hat{\mathbf{d}}$-vector can point in any direction.  % but in the same direction on \(\Gamma\) and \(K\): $\mathbf{d}_{\Gamma}\left(\mathbf{p}\right)=D_{\Gamma}^{(t)}\cos3\theta\,\hat{\mathbf{d}}$
% is the d-vector of 
%on the $\Gamma$ pocket gap, %whereas
%and $\mathbf{d}_{K}\left(\mathbf{p}\right)=D_{K}^{(t)}\,\hat{\mathbf{d}}$.} %gap.
Unlike typical triplet gaps,
here %$\mathbf{d}_{K}\left(\mathbf{p}\right)$ 
$\Delta_{K}\left(\mathbf{p}\right)$ is momentum independent on the $K$-pocket, since for triplet states we have $\Delta_{-K}\left(\mathbf{p}\right) =- \Delta_{K}\left(\mathbf{p}\right)$. 
In order for $\Delta_{\Gamma}\left(\mathbf{p}\right)$ to be non-zero,
we include the sub-leading momentum-direction-dependent
interactions (\ref{tripInt}), which do not contribute significantly to the pairing
instability. %, must be included.
While here our focus is on SC due to purely electronic interactions,
the SC states obtained above are not necessarily inconsistent with
electron-phonon interactions, which are expected to promote intra-pocket attraction, thus reducing the amplitude \textendash{} or even changing the sign \textendash{}
of the $g_{1}$, $g_{2}$ terms. 
\subsection{Density Wave Instabilities}

Next, we show that the interactions $g_1, ... g_4$ do not lead to density-wave particle-hole instabilities in the spin and charge channels within logarithmic accuracy. This means that any instability in these channels requires a threshold value for these interactions, and is thus unlikely to be driven by the low-energy fermions. The vertices associated with the particle-hole instabilities are
\begin{equation}
\Delta^{(\mu DW)}_{\eta,\eta'} d^\dagger_{\eta \alpha}\sigma^\mu_{\alpha\beta} d_{\eta'\beta}
\end{equation}
with \(\mu=0\) corresponding to CDW and \(\mu=x,y,z\) to SDW order parameters (here we are ignoring the small momentum dependence). Because the pockets at $K$ and $\Gamma$ are both hole pockets, the spin and charge density-wave channels completely decouple from the SC channels. Explicitly, while the particle-particle bubble
\begin{equation}
\Pi_\eta=\int \frac{\tanh{\frac{\epsilon_\eta}{2T}}}{2\epsilon_\eta}\frac{d^2Q}{(2\pi)^3}
\end{equation}
has a logarithmic divergence when integrated over all momenta, the particle-hole bubble does not:
\begin{eqnarray}\label{Chi}
\chi_{\eta}&=T\sum_\omega\int G^{(0)}_\eta(i\omega,\mathbf{Q})G^{(0)}_{-\eta}(i\omega,-\mathbf{Q})\frac{d^2Q}{(2\pi)^3}\nonumber\\
&=-\int\frac{\sech{\frac{\epsilon_\eta}{2T}}}{4T}\frac{d^2Q}{(2\pi)^2}
\end{eqnarray}
Consequently, any particle-hole channel is subleading to the SC channel within weak-coupling. For example,
\begin{equation}
\delta \Delta^{(CDW)}_{\Gamma,\Gamma}=2\chi_{\Gamma}g_1\Delta^{(CDW)}_{\Gamma,\Gamma}+\chi_{K}\frac{g_6+g_7}{2}\Delta^{(CDW)}_{K,K}
\end{equation}
There are additional CDW and SDW vertices, but all the equations have \(\chi_{\eta}\) in them and thus the flows are all exponentially suppressed by a factor of \(\sech\frac{\Lambda}{2T}\approx 2e^{-\frac{\Lambda}{2T}}\) as a result. This agrees with the analysis of Ref. \cite{Mengxing18}, which only found leading particle-hole instabilities because the $\Gamma$ pocket was electron-like and nested with the $K$ pocket. Note that an incommensurate CDW has in fact been observed in monolayer NbSe\(_2\) \cite{1HCDW}, as well as in bulk. While its origin remains controversial, the evidence supports a scenario in which the CDW is not a consequence of Fermi surface nesting, consistent with our calculation \cite{JohannesCDW06, CalandraCDW09, Borisenko, RafaelCDW}.

\section{Superconductivity in the presence of %Rashba 
SOC and magnetic field}\label{IV}

The parquet RG treatment of the previous section shows that superconductivity is the only instability generated by the interactions in Eq. (\ref{H_int}) within weak-coupling. If the superconducting transition temperature was larger than the Fermi energy, then the superconducting problem would have been essentially solved. However, since $T_c \ll E_F$ in NbSe$_2$, one needs to proceed to energy scales below the Fermi energy, where Ising and Rashba SOC and the magnetic field become relevant. Since superconductivity is the leading instability and decouples from other channels below the Fermi energy, it is sufficient to consider a simpler mean-field approach to include the additional terms in the Hamiltonian that were neglected in analysis above. 

\subsection{Mean-Field Gap Equation in the Presence of SOC and Magnetic Field}\label{SecIVA}

In order to determine the appropriate BCS gap equation, we first 
project the interactions onto the spin-split Fermi surfaces, by using the transformation (\ref{Cbasis}) between eigenstates of spin and eigenstates of the non-interacting Hamiltonian and restricting the gap functions to involve pairs on the same Fermi surface only.  Throughout the phase diagrams shown in Fig. \ref{PhaseFig}, the Fermi surfaces are qualitatively similar to those shown in Fig. \ref{FSFig}, with the important caveat that in the presence of both Rashba SOC and magnetic field inversion symmetry is broken, as we discuss in more detail below.  Our analysis assumes that the minimum splitting between Fermi surfaces (with an associated energy scale on the order of the magnetic field \(b\) and/or Rashba SOC \(\alpha_Rp_F\)) is large compared to the superconducting pairing strength, whose energy scale is on the order of the superconducting gap function. In particular, close to the phase transition at which the gap function vanishes this is valid almost everywhere in our phase diagram.  Note, however, that for $\alpha_R = b = 0$, or for $\alpha_R = b$ with ${\bf B}$ along a $\Gamma -K$ line, the inner and outer Fermi surfaces at the $\Gamma$ pocket touch along the $\Gamma-M$ lines, and our approach is insufficient to resolve the gap in the immediate vicinity of these loci even very close to $T_c$.    

After the
projection, the interactions generally acquire a dependence on the momentum direction. Explicitly, this gives:
\begin{align}
H_{\text{Int}}= & \tilde{V}_{\Gamma,\Gamma}^{\tau,\tau'}c_{\Gamma\tau}^{\dagger}c_{\Gamma\tau}^{\dagger}c_{\Gamma\tau'}c_{\Gamma\tau'}+\nonumber \\
 & \tilde{V}_{\pm K,\pm K}^{\tau,\tau'}c_{\pm K\tau}^{\dagger}c_{\mp K\tau}^{\dagger}c_{\pm K\tau'}c_{\mp K\tau'}+\nonumber \\
 & \tilde{V}_{\pm K,\mp K}^{\tau,\tau'}c_{\pm K\tau}^{\dagger}c_{\mp K\tau}^{\dagger}c_{\mp K\tau'}c_{\pm K\tau'}+\nonumber \\
 & \tilde{V}_{\Gamma\pm K}^{\tau,\tau'}c_{\pm K\tau}^{\dagger}c_{\mp K\tau}^{\dagger}c_{\Gamma\tau'}c_{\Gamma\tau'}
\end{align}
with %(repeated indices are summed implicitly in the expression below):
%\begin{multline}
%\tilde{V}_{\eta,\eta'}^{\tau,\tau'}(\mathbf{p},\mathbf{k})=\\
% U_{\eta\tau}^{\alpha}(\mathbf{p})U_{-\eta\tau}^{\beta}(-\mathbf{p})U_{\eta'\tau'}^{\alpha'*}(\mathbf{k})U_{-\eta'\tau'}^{\beta'*}(-\mathbf{k})V_{\eta,\eta'}^{\alpha\beta;\alpha'\beta'}(\mathbf{p};\mathbf{k})
%\end{multline}
%{\color{blue}Conveniently, this has the form}
\begin{equation} \label{Eq:VQs}
\tilde{V}_{\eta,\eta'}^{\tau,\tau'}(\mathbf{p},\mathbf{k})=\frac{1}{2}\sum_{\mu=0,x,y,z}g_{\eta,\eta'}^{(\mu)}Q_{\eta\tau}^{(\mu)}(\mathbf{p})Q_{\eta'\tau'}^{(\mu)*}(\mathbf{k})
\end{equation}
%\begin{align}
%W_{\eta\tau}^{(i)}(\mathbf{p})=\sum_{\alpha\beta}(i\sigma^{i}i\sigma^{y})_{\alpha\beta}U_{\eta\tau}^{\alpha}(\mathbf{p})U_{-\eta\tau}^{\beta}(-\mathbf{p})
%\end{align}
(\(\eta, \eta'=\Gamma,\pm K\)).  Here terms on the right-hand side with \(\mu=0\) are projections of singlet interactions in (\ref{Eq:Ints_SpinChannel}), while those with \(\mu=j\equiv x,y,z\) are projections of the three triplet interactions respectively, into the relevant Fermi surface. Since spin is not conserved in the presence of SOC and/or magnetic field, these three spin polarizations are no longer equivalent.  Explicitly, 
\begin{eqnarray} \label{Eq:GapFuns}
Q_{\eta\tau}^{(0)}(\mathbf{p}) & = &% W_{\eta\tau}^{(0)}(\mathbf{p})=
\sum_{\alpha\beta}(i\sigma^{y})_{\alpha\beta}U_{\eta\tau}^{\alpha}(\mathbf{p})U_{-\eta\tau}^{\beta}(-\mathbf{p})   \\
Q_{\pm K\tau}^{(j)}(\mathbf{p}) & = & % \pm W_{\pm K\tau}^{(i)}(\mathbf{p})
\pm \sum_{\alpha\beta}(i\sigma^{j}i\sigma^{y})_{\alpha\beta}U_{K \tau}^{\alpha}(\mathbf{p})U_{-K\tau}^{\beta}(-\mathbf{p}) \nonumber  \\
Q_{\Gamma\tau}^{(j)}(\mathbf{p}) & = & \sqrt{2}\cos(3\theta_{{\bf p}}) %W_{\Gamma\tau}^{(i)}(\mathbf{p}) 
\sum_{\alpha\beta}(i\sigma^{j}i\sigma^{y})_{\alpha\beta}U_{\Gamma\tau}^{\alpha}(\mathbf{p})U_{\Gamma\tau}^{\beta}(-\mathbf{p})   \nonumber  \ .
\end{eqnarray}
 There is a phase ambiguity in the definitions of \(Q_{\eta\tau}^{(\mu)}(\mathbf{p})\); the additional factors of \(i\) in the last two expressions  are chosen to simplify the gap equation below. Here $g_{\eta,\eta'}^{(\mu)}$ are constants independent of $\mathbf{p}$ and
$\mathbf{k}$ (with \(j=x,y,z\)):
\begin{eqnarray}
&g^{(0)}_{\Gamma,\Gamma}& =g_1 \qquad g^{(j)}_{\Gamma,\Gamma} =g_1^t\nonumber\\
&g^{(0)}_{\Gamma,\pm K}&=g^{(0)}_{\pm K,\Gamma} = g_4\nonumber\\
&g^{(j)}_{\Gamma,\pm K}&=g^{(j)}_{\pm K,\Gamma} =g_4^t\nonumber\\
&g^{(0)}_{\pm K,\pm K}&=g^{(0)}_{\pm K,\mp K} = \frac{g_2+g_3}{2} \nonumber\\
& g^{(j)}_{\pm K,\pm K}&=g^{(j)}_{\pm K,\mp K} =\frac{g_2-g_3}{2}
\end{eqnarray}
where $g_1,... g_4$ are the values of the couplings at the end of the RG procedure described above.

For uniform SC
(i.e. with zero center-of-mass momentum), the paired electrons are either both from inner
pockets or both from outer pockets, with corresponding %Performing a Hubbard-Stratonovich transformation introduces the 
gap functions 
%\begin{eqnarray}
%\Delta_{\eta\tau}({\bf p}) & \propto & \langle c_{\eta,{\bf p}\tau}c_{-\eta,-{\bf p}\tau}\rangle
%\end{eqnarray}
\begin{eqnarray}
\Delta_{\Gamma \tau}({\bf p}) & \propto & \langle c_{\Gamma,{\bf p}\tau}c_{\Gamma,-{\bf p}\tau}\rangle\nonumber \\
\Delta_{\pm K \tau}({\bf p}) & \propto &  \langle c_{ \pm K,{\bf p}\tau}c_{\mp K,-{\bf p}\tau}\rangle
\end{eqnarray}
where %$\epsilon = \pm 1$ and
the momentum ${\bf p}$ is measured with respect to the center of the relevant Fermi pocket. % at \(\eta\).
The gaps are diagonal in the index $\tau$, as we assume there is no pairing between inner and outer Fermi surfaces. %(as mentioned above, this is valid sufficiently close to the phase transition).
Note that particle-hole symmetry imposes $\Delta_{-K\tau}\left(\mathbf{p}\right)=-\Delta_{K\tau}\left(-\mathbf{p}\right)$, so these are not two separate order parameters.
The new self-consistent gap equation, analogous to Eq. (\ref{Eq:VertexFlow}), is (see Appendix \ref{A}):
\begin{equation}
\Delta_{\eta\tau}(\mathbf{p})=\sum_{\eta',\tau'}\oint\Pi_{\eta'\tau'}(\theta_{\eta',\mathbf{k}})\tilde{V}_{\eta,\eta'}^{\tau,\tau'}(\mathbf{p};\mathbf{k})\Delta_{\eta'\tau'}(\mathbf{k})\frac{d\theta_{\eta',\mathbf{k}}}{2\pi}\label{Eq:Gap}
\end{equation}
After integrating over momenta, the (angle resolved) particle-particle pairing susceptibility $\Pi_{\eta\tau}(\theta_{\eta,\mathbf{p}})$ becomes: %removed equation reference; these are not the same as in    (\ref{Eq:Pi})
\begin{equation}\label{Pi}
\Pi_{\eta\tau}(\theta_{\eta,\mathbf{p}})=-\sum_{p}\frac{\tanh\left(\frac{\beta \xi_{\eta\tau}(\mathbf{p})}{2}\right)+\tanh\left(\frac{\beta \xi_{-\eta\tau}(-\mathbf{p})}{2}\right)}{\xi_{\eta\tau}(\mathbf{p})+\xi_{-\eta\tau}(-\mathbf{p})}
\end{equation}
where %$\xi_{\eta\tau}(\mathbf{p})$ is given in Eq. (\ref{Eq:Xidef}), and 
\(\theta_{\eta,\mathbf{p}}\) is the angle along the \(\tau\) Fermi surface relative to the center of the pocket $\eta$ (below we simply use \(\theta\) when this is clear from context). Assuming that the Fermi surface is inversion symmetric, % (i.e. \(\xi_{A\eta\tau} = 0\))
 we find
\begin{equation}\label{PiXiA0}
\Pi_{\eta\tau}(\theta_{\eta,\mathbf{p}})=-N_{\eta\tau}\ln\frac{1.13\Lambda}{T_{c}}
\end{equation}
with $N_{\eta\tau}$ the density of states at the inner or outer Fermi surface at the \(\eta\) pocket and \(\Lambda\) is the cut-off energy. Although the Fermi surface is only inversion symmetric when \(\alpha_R=0\) or \(b=0\),  the expression (\ref{PiXiA0}) is approximately correct when either is sufficiently small; this issue will be discussed in more detail in Sec. \ref{SecIVC}.

Plugging Eq. (\ref{Eq:VQs}) into Eq. (\ref{Eq:Gap}), we find
\begin{equation}\label{GapFnc}
\Delta_{\eta\tau}(\mathbf{k})=\sum_{\mu}D_{\eta\tau}^{(\mu)}Q_{\eta\tau}^{(\mu)}(\mathbf{p})
\end{equation}
where $\mathbf{D}_{\eta\tau}=(D_{\eta\tau}^{(0)},D_{\eta\tau}^{(x)},D_{\eta\tau}^{(y)},D_{\eta\tau}^{(z)})$ are momentum independent gap coefficients.
%, making the $D_{\eta\tau}^{(\mu)}$ coefficients real when the density of states are equal on inner and outer Fermi surfaces.
The structure of Eq. (\ref{Eq:Gap}) implies that we can take $D_{\eta\tau}^{(\mu)}=D_{\eta-\tau}^{(\mu)} \equiv D_{\eta}^{(\mu)}$,
and we thus drop the $\tau$ index on $D$ hereafter.  Moreover, particle-hole
symmetry enforces $D_{K}^{(\mu)}=D_{-K}^{(\mu)}$, consistent with the fact that $\Delta_{-K\tau}\left(\mathbf{p}\right)=-\Delta_{K\tau}\left(-\mathbf{p}\right)$.
Plugging the form (\ref{GapFnc}) back into the gap equation (\ref{Eq:Gap})
yields the reduced gap equations 
\begin{eqnarray}\label{Eq:GapFinds}
D_{\Gamma}^{(0)}= & \sum_{\mu}\left(g_{1}f_{(\mu)}^{(0)\Gamma}D_{\Gamma}^{(\mu)}+2g_{4}f_{(\mu)}^{(0)K}D_{K}^{(\mu)}\right)\label{Eq:Gapsols}\\
D_{K}^{(0)}= & \sum_{\mu}\left(g_{4}f_{(\mu)}^{(0)\Gamma}D_{\Gamma}^{(\mu)}+(g_{2}+g_{3})f_{(\mu)}^{(0)K}D_{K}^{(\mu)}\right)\nonumber \\
D_{\Gamma}^{(j)}= & \sum_{\mu}\left(g_{1}^{t}f_{(\mu)}^{(j)\Gamma}D_{\Gamma}^{(\mu)}+2g_{4}^{t}f_{(\mu)}^{(j)K}D_{K}^{(\mu)}\right)\nonumber \\
D_{K}^{(j)}= & \sum_{\mu}\left(g_{4}^{t}f_{(\mu)}^{(j)\Gamma}D_{\Gamma}^{(\mu)}+(g_{2}-g_{3})f_{(\mu)}^{(j)K}D_{K}^{(\mu)}\right)\nonumber 
\end{eqnarray}
where $j=x, y, z$ and the form factors $f_{(\mu)}^{(\mu')\eta}$ are
given by
\begin{align}\label{Eq:FormFactors}
f_{(\mu)}^{(\mu') \eta}= &\frac{1}{2} \oint\sum_{\tau}\Pi_{\eta\tau}Q_{\eta\tau}^{(\mu)*}Q_{\eta\tau}^{(\mu')}\frac{d\theta_{\eta,\mathbf{k}}}{2\pi}
\end{align}
%{\color{blue}See Appendix \ref{B} for more explicit expressions.}  
In the presence of SOC  and magnetic field none of the channels $(\mu = 0, x,y,z)$ decouple in general.  Eqs. (\ref{Eq:GapFinds}) thus can be viewed as an $8\times8$
matrix equation,
 leading to 8 possible superconducting solutions,
of which we choose the one with the highest $T_{c}$.  %the singlet and triplet channels do not decouple, and the superconducting gaps are neither spin singlet nor spin triplet. 
The choice of phases in Eq. (\ref{Eq:GapFuns}) was made to make the form factors \(f_{(\mu)}^{(\mu') \eta}\) real when the densities of state on inner and outer Fermi surfaces are equal %(see Appendix \ref{B}), 
in which case the coefficients \(D_{\eta}^{(\mu)}\) can also be taken to be real. %(corresponding to \(s+if\) mixing in the language of \cite{Khodas19})}.  {\fb Make the last comment a footnote?} The solutions can be found analytically when either the magnetic field or Rashba SOC is absent (see Appendix \ref{B} for explicit expressions), but otherwise the equations have to be solved numerically.

\subsection{Phase Diagrams}\label{SecIVB}

%{\color{cyan}We now go over the solutions of the gap equation, Eq.'s (\ref{GapFnc}-\ref{Eq:FormFactors}) and describe the corresponding phase diagrams shown in Fig. \ref{PhaseFig}.}
%{\color{blue}As spin is not conserved in the presence of SOC, it is no longer possible to talk about spin-singlet or spin-triplet pairing. However,} 

To study the possible superconducting phases, it is useful to define singlet and triplet instability regimes by considering the limit of no SOC and magnetic field. % {\color{cyan}and to consider two separate phase diagrams for each regime, shown in Fig. \ref{PhaseFig}(a) and (b) respectively. In particular,} 
We define a dominant singlet (dominant triplet) instability to occur when the largest eigenvalue of the gap equation %(\ref{Eq:Gapsols})
is for the spin singlet (spin triplet) gap with SOC and magnetic field set to 0. %Note: additional qualification is not redundant
The transition temperature for each channel (when the corresponding eigenvalue of the gap equation equals $1$) is determined by the couplings $g_1,...g_4$ via %. This yields 
\begin{equation}
T_c^{(a)} =1.13 \Lambda e^{-1\slash{\gamma^{(a+)}}} \ ,
\end{equation}  where %$N$ is the DOS of all bands (assumed to be equal), 
$\Lambda$ is the upper energy cutoff, and \(\gamma^{(a+)}\) are identical to the couplings obtained in the RG analysis in Eq. (\ref{Eq:gamma}) (\(a=s,t\) for singlet and triplet respectively).
%We use this definition of the singlet and triplet instability regimes depend only on the coupling constants, and applies even in the presence of SOC and magnetic field. 
 Fig. \ref{PhaseFig} shows the phase diagrams corresponding to singlet (a) and triplet (b) instability regimes as a function of Rashba SOC and in-plane magnetic field.  We emphasize that the resulting SC states themselves are always a mixture of singlet and triplet Cooper pairs.

As seen in Eqs. (\ref{gamma0}) and (\ref{gammat}), with equal DOS in all bands, for repulsive interactions the singlet instability dominates for large $g_4$, while the triplet instability dominates for large $g_3$. For concreteness, in this section we take 
\begin{equation}\label{ParamG}
g_{2}=1.2g_1,\quad g_{4}=2g_1,\quad g_{1}^{t}=0.2g_1,\quad g_{4}^{t}=0.1g_1
\end{equation}
with \(g_{3}=1.05g_1\) (\(4.2g_1\)) to produce a singlet (triplet) instability.

To describe the phase diagrams in Fig. \ref{PhaseFig}, 
it is useful to classify the solutions to the gap equations (\ref{GapFnc}) by the irreducible representations (irreps) of the relevant point group. In the absence of Rashba SOC %\(\alpha_Rp_F\) 
and magnetic field, %\(\mathbf{b}=g_L\mu_B\mathbf{B}\), 
 the point group of 1H-NbSe\(_2\) is \(D_{3h}\) \footnote{The relevant functional forms for the gap in each irrep can be found in the spin basis in Ref. \cite{Khodas19}; these are related to our gaps by the basis transformation (\ref{Cbasis}).}.  We find that %the spin-singlet gap and the spin-triplet gap with d-vector along \(z\) in Eq. (\ref{Eq:Gap}) 
%transform according to the \(A_1'\) irrep of \(D_{3h}\), while the spin-triplet gaps with the d-vector aligned in-plane in either the \(x\) or \(y\) directions transform as two components of the two dimensional \(E''\) irrep of \(D_{3h}\). Their projections transform according to the same irreps: 
the \(\mu=0, z\) terms on the right-hand side of Eq. (\ref{GapFnc}) both belong to the \(A_1'\) irrep of \(D_{3h}\), indicating that the singlet and $z$-polarized spin-triplet gaps are mixed in the presence of Ising SOC. In our model, this mixing is proportional to the difference between the densities of states \(N_{\eta\tau}\) on the inner $(\tau = -1)$ and outer $(\tau = 1)$ Fermi surfaces. %, i.e. \(f_{(z)}^{(0) \eta}\propto N_{\eta1}-N_{\eta-1}\) in Eq. (\ref{Eq:FormFactors}) (see Appendix \ref{B} for explicit form of the form factors).  
Since these densities of states are not expected to differ significantly in two dimensions, this mixing is weak in our model. 
The remaining components \(\mu=x, y\) of the triplet gap transform as the two dimensional \(E''\) irrep of \(D_{3h}\).   
We find that, in the absence of both Rashba SOC and magnetic field, the highest $T_c$ corresponds to the $A_{1}'$ irrep in our model. %, so that \((x)\) and \((y)\) terms in the gap equation vanish in this limit (see Appendix \ref{B} for the explicit form of the corresponding gap functions).  
%  The main distinction between the singlet and triplet instability regimes, therefore, comes rather from the behavior of the \((x)\) and \((y)\) terms favored by the triplet instability, as we will see below.

Rashba SOC transforms as the \(A_2''\) irrep of \(D_{3h}\). This lowers the point group to \(C_{3v}\), but does not mix the \(A_1'\) and \(E''\) gaps. %\(D_{3h}\) and irreps \(A_1'\) and \(E''\) of \(D_{3h}\) descend to \(A_1\) and \(E\) irreps of \(C_{3v}\) respectively. This means that adding Rashba SOC does not introduce new mixing terms in the gap equation.
In contrast, the in-plane magnetic field \(\mathbf{b}\) transforms according to the \(E''\) irrep of \(D_{3h}\). As a result, 
it mixes the \(E''\) gap with the \(A_1'\) one  \cite{Khodas19}. %(resulting in so-called \(s+if\) mixing \cite{Khodas19}; see Appendix \ref{B} for details)
Thus, in the presence of an in-plane magnetic field, all \(\mu\) terms in Eq. (\ref{GapFnc}) are mixed.

The electronic spectrum in the superconducting phase is obtained by diagonalizing the Bogolyubov-de Gennes (BdG) Hamiltonian:
\begin{equation}
H=\frac{1}{2}\sum_{\mathbf{p}\eta\tau}\Psi_{\mathbf{p}\eta\tau}^\dagger\mathcal{H}_{\eta\tau}(\mathbf{p})\Psi_{\mathbf{p}\eta\tau}
\end{equation}
where \(\Psi_{\eta\tau}(\mathbf{p})=(c_{\eta,\mathbf{p}\tau},c^\dagger_{-\eta,\mathbf{-p}\tau})^T\) and 
\begin{equation}\label{BdG}
\mathcal{H}_{\eta\tau}(\mathbf{p})=\left(\begin{array}{cc}
\xi_{\eta\tau}(\mathbf{p}) & \Delta_{\eta\tau}(\mathbf{p}) \\
\Delta_{\eta\tau}^*(\mathbf{p}) & -\xi_{-\eta\tau}(\mathbf{-p})
\end{array}\right)
\end{equation}
with \(\xi_{\eta\tau}(\mathbf{p})\)  given in Eq. (\ref{Eq:Xidef}). Note that when time reversal symmetry (TRS) is broken, \(\xi_{-\eta\tau}(\mathbf{-p})\neq\xi_{\eta\tau}(\mathbf{p})\) in general. The BdG spectra are given by $E_{\eta\tau}(\mathbf{p}),-E_{-\eta\tau}(-\mathbf{p})$, with:
\begin{equation}\label{Eq:BdGspec}
E_{\eta\tau}(\mathbf{p})=\xi_{A\eta\tau}(\mathbf{p})+\sqrt{\xi_{S\eta\tau}(\mathbf{p})^2+|\Delta_{\eta\tau}(\mathbf{p})|^2}
\end{equation}
where
\begin{eqnarray}\label{Eq:Xi}
\xi_{S\eta\tau}(\mathbf{p})=\frac{\xi_{\eta\tau}(\mathbf{p})+\xi_{-\eta\tau}(-\mathbf{p})}{2}\label{Eq:XiS}\\
\xi_{A\eta\tau}(\mathbf{p})=\frac{\xi_{\eta\tau}(\mathbf{p})-\xi_{-\eta\tau}(-\mathbf{p})}{2}\label{Eq:Xi}
\end{eqnarray}
Clearly, nodes only occur if both \(\xi_{S\eta\tau}\) and \(\Delta_{\eta\tau}\) vanish simultaneously. Note that, in our case, the Fermi surface is symmetric under momentum inversion, \(\mathbf{p}\rightarrow-\mathbf{p}\), only when either \(\alpha_R\) or \(b\) vanish. In this case, \(\xi_{A\eta\tau}=0\) and \(\xi_{S\eta\tau}=\xi_{\eta\tau}\).  
%In presence of momentum inversion symmetry, \(\xi_{S\eta\tau}=\xi_{\eta\tau}=0\) simply defines the Fermi surface, so the nodes occur whenever the gap function vanishes along the Fermi surface. The BdG spectra are plotted in Fig. \ref{GapFig}.

\begin{figure}
\centering \includegraphics[width=0.99\columnwidth]{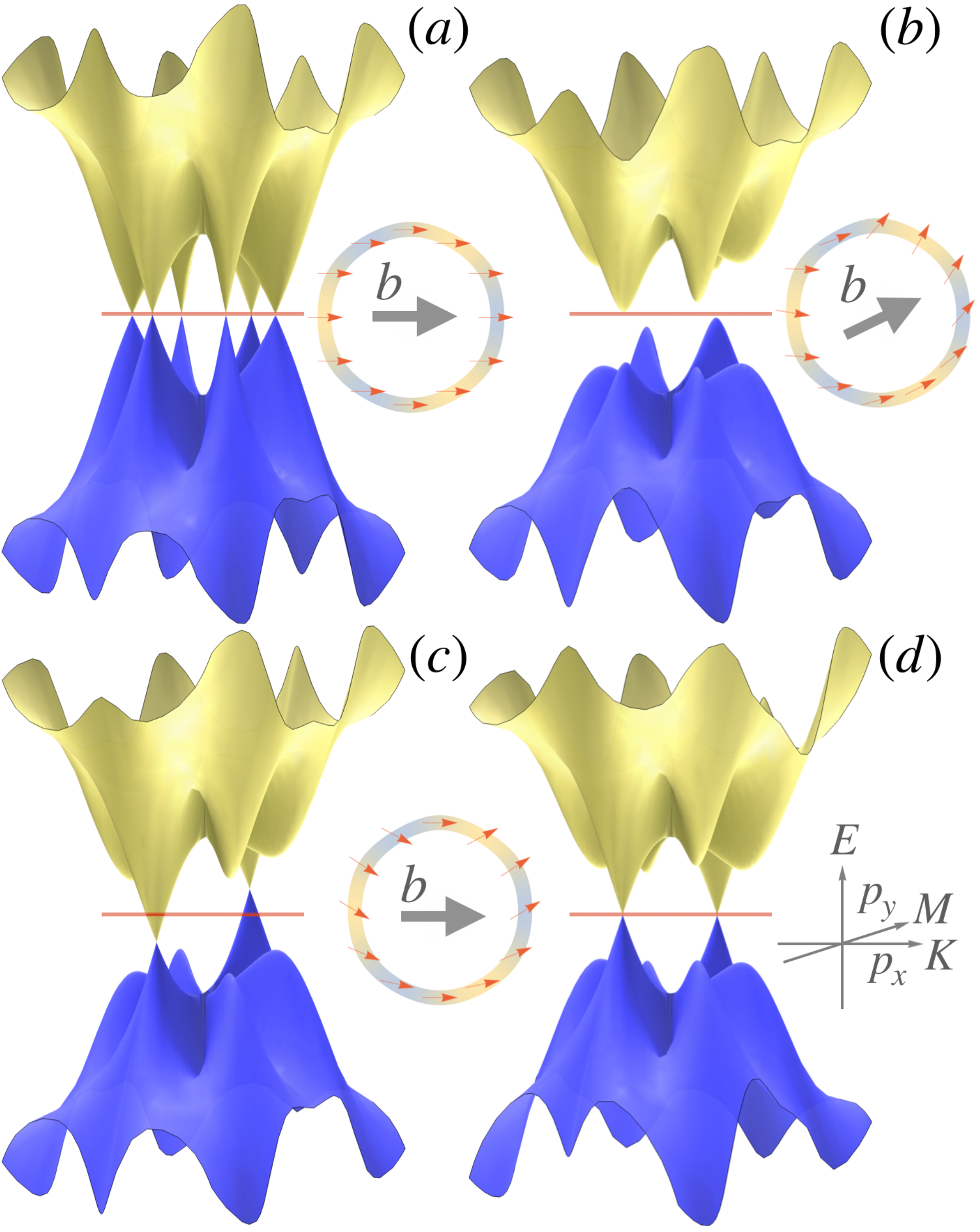}
\\
\caption{Superconducting excitation spectrum, Eq. (\ref{Eq:BdGspec}), for the inner %{\color{green} \bf{The main text says it's the outer Fermi surface. Which one is it?}}
Fermi surface at $\Gamma$ in the presence of an in-plane magnetic field and without (panel a) or with (panels b, c, d) Rashba SOC. In panel b (panels c and d), $\mathbf{B}$ is aligned 
along \(\vartheta=\pi/7\) (\(\vartheta=0\)) from the $\Gamma$-$K$ lines; panel a is the same for any field direction. In panel c (panel d), the Cooper pair has zero (non-zero) center-of-mass momentum. Insets show the resulting spin textures along the normal-state Fermi surfaces for the corresponding field directions, with colors as in Fig. \ref{FSFig} and arrows indicating in-plane spin components.  We used the normal state dispersion from Eq. (\ref{Eq:H0z}) with the same parameter values given in Eq. (\ref{ParamFS}), and took \(b=5\beta_I\), as well as \(\alpha_R=2\beta_I\) in panels b-d. We set \(D_{\Gamma,-1}^{(0)}=20\beta_I\) and \(D_{\Gamma,-1}^{(z)}=20\beta_I\) in Eq. (\ref{GapFnc}), and \(p_\mathrm{{shift}}=-0.3p_F\) in panel d.}
\label{GapFig} % \end{center}
\end{figure}

We are now in position to describe the
SC phase diagrams of Fig. \ref{PhaseFig}, obtained for a fixed value of the Ising SOC and varying the magnitude of the magnetic field $b$ and Rashba SOC $\alpha_{R}$.  
In all cases, the gap at the $\pm K$ pockets is
nearly isotropic, so we focus on the \(\Gamma\) pocket. 
We first analyze the phase diagram of Fig. \ref{PhaseFig}(a), where the dominant $g_4$ interaction gives the singlet
extended $s$-wave state in the limit of vanishing SOC and magnetic field.
Along the $b=0$ axis, the main effect of increasing
the Rashba SOC $\alpha_{R}$ is to change the anisotropy of $\Delta_{\Gamma\tau}\left(\mathbf{p}\right)$,
due to the small admixture with the $\mu = z$ nodal triplet gap.
%(As noted above this admixture is symmetry-allowed in principle, but strongly suppressed in practise, with only Ising SOC).  
 Importantly, no phase transition happens along this
axis, as the dominant instability is always in the $A_1'$ irrep of the original $D_{3h}$ point group.   In contrast, along the $\alpha_{R}=0$ axis, a phase transition
takes place to a nodal topological SC state for $b=b_{P}$, where
$b_{P}\approx\Delta_{\Gamma1}$ corresponds roughly to the Pauli-limiting
field \cite{Chandrasekhar,Clogston}. Because our assumption of well-separated Fermi surfaces is not valid for fields smaller than $b_P$, our analysis is not sufficient to determine the phase boundary quantitatively, but we show it qualitatively in Fig. \ref{PhaseFig}(a). This phase transition, and the
topological character of the resulting nodal SC state, were previously
predicted in Ref. \cite{LawMay16} and can be understood as a consequence
of the vanishing of the Ising SOC along the six $\Gamma$-$M$ directions,
where the SC gap vanishes and 12 nodes (6 for each $\Gamma$ Fermi surface) appear due to spins
aligning with the magnetic field. This gap structure is shown in the BdG spectrum of the inner $\Gamma$ Fermi surface, displayed in Fig. \ref{GapFig}(a) together with the spin texture of the normal-state Fermi surface.

\begin{figure}%[htp]
\centering \includegraphics[width=0.99\columnwidth]{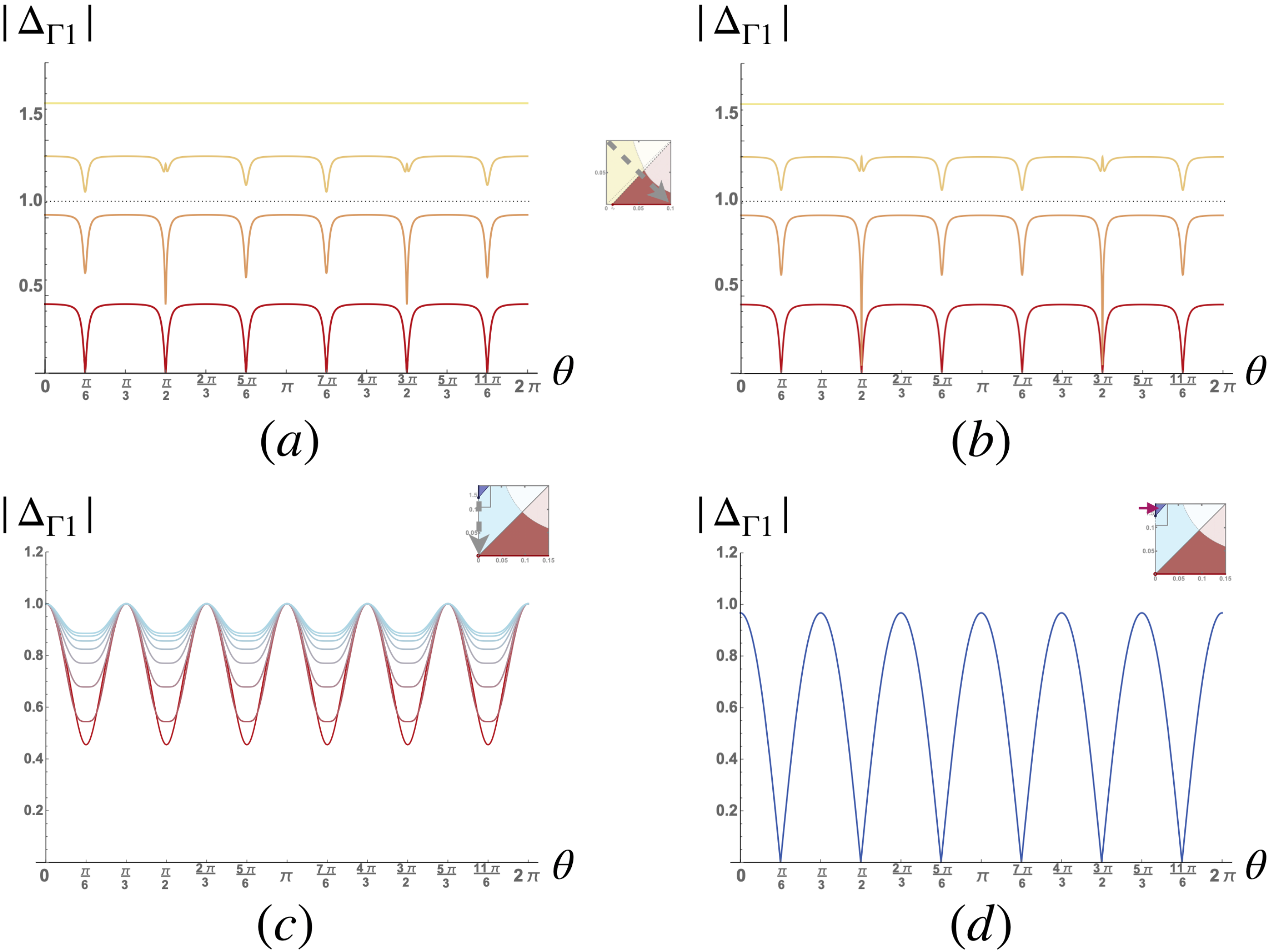}
\caption{Superconducting gap $\Delta_{\Gamma1}$ in Eq. (\ref{GapFnc}) at the outer $\Gamma$ pocket,
as a function of the angle $\theta$ along the Fermi surface with respect to the $\Gamma$-$K$
direction, in various regions of the phase diagrams of Fig. \ref{PhaseFig}.
Panels (a) and (b) correspond to the cuts across the phase diagram of Fig. \ref{PhaseFig}(a) shown
in the insets, with a magnetic field away from the $\Gamma$-$K$
direction by an angle \(\vartheta=2\pi/25\) (panel (a)) and along  the $\Gamma$-$K$ direction (panel (b)).
Panels (c) and (d) correspond to cuts along the $b=0$ axis of the
phase diagram of Fig.  \ref{PhaseFig}(b), outside and inside the chiral
SC phase, respectively (see insets). Note that the gap amplitudes
have been rescaled for clarity, since they are not fixed by the linearized
gap equations. We used coupling constants given in Eq. (\ref{ParamG}) and %we took \(m=1.5p_F^2\slash\beta_I\), \(\mu=-5\beta_I\) and \(\lambda p_F^3=0.6 \beta_I\) 
parameters for the non-interacting Hamiltonian as in Eq. (\ref{ParamFS}).  We also took
the inner and outer densities of states to differ by ten percent to ensure that the symmetry allowed mixings between the singlet and triplet channels are present in our solutions.}
\label{Gaps} % \end{center}
 \end{figure}

Turning on the Rashba SOC introduces a second spin-orbit energy scale that does not vanish along
the $\Gamma$-$M$ directions. % {\color{cyan}and breaks the time reversal-like symmetry protecting the nodes}. 
As a result, generally even an infinitesimal
Rashba SOC lifts the nodes and destroys the topological character
of this state, as shown by the fully gapped BdG spectrum in Fig. \ref{GapFig}(b).
The only exception is when $\mathbf{b}$ is aligned along one of the $\Gamma$-$K$ directions: in this case, as we discuss
in detail in Sec. \ref{V}, the system has a mirror symmetry.
For $\alpha_{R}p_{F}<b$, this symmetry forces spins along the $\Gamma$-$M$ line perpendicular
to $\mathbf{b}$ to align (anti-align) with the magnetic field on the inner (outer) \(\Gamma\) pocket, as shown in the inset of Fig. \ref{GapFig}(c).
As a consequence, the gap vanishes along the line perpendicular
to $\mathbf{b}$, as displayed by the BdG spectrum of Fig. \ref{GapFig}(c). Therefore, two pairs of nodes originally present on this line are protected, whereas the remaining eight nodes are gapped, resulting in a \emph{crystalline} gapless topological SC state.  Because the Fermi surfaces are no longer symmetric under momentum inversion (i.e. \(\xi_{A\eta\tau}\neq0\)), these protected nodes are generally shifted away from the Fermi level,  resulting in the Bogolyubov Fermi surfaces shown in Fig. \ref{GapFig}(c). As we discuss in the next section, however, the nodes can move back to the Fermi level if the Cooper pair acquires a finite center-of-mass momentum, which is expected to happen for large enough $b$ and $\alpha_R$ (Fig. \ref{GapFig}(d)).
For  \(\alpha_Rp_F>b\), however, the pair of nodes on the inner and outer Fermi surfaces merge and the superconducting state becomes
fully gapped. While we could not precisely locate this phase boundary, it is expected to interpolate between $b = \alpha_Rp_F$ for large values of $\alpha_R$ to $b = b_P$ for $\alpha_R = 0$, as shown by the dashed line in Fig. \ref{PhaseFig}(a). The evolution of the gap at the outer $\Gamma$ Fermi surface across this transition is shown in Figs. \ref{Gaps}(a) and (b). In panel (a), the magnetic field is applied along a direction that does not coincide with the $\Gamma-K$ direction. As a result, the nodal superconducting state only exists for $\alpha_R = 0$ (red curve). In contrast, when the magnetic field is applied along the $\Gamma-K$ direction (panel (b)), two nodes persist even when $\alpha_R \neq 0$ (dark orange curve).

We now turn to the case of dominant triplet instability shown in Fig. \ref{PhaseFig}(b), obtained for a dominant $g_3$ interaction.   
As in the singlet regime, we observe a nodal topological superconductor for $\alpha_R= 0$. In the triplet regime, however, the superconducting gap on the $\Gamma$ pocket is nodal along the entire $\alpha_R = 0$ line (except for very small magnetic fields, where the small difference in density of states on the inner and outer Fermi surfaces can open a gap); hence the nodal topological SC state occurs for all values of $b$, as there is no Pauli limit in this case \cite{Chandrasekhar,Clogston}.  
Similarly, the transition into the crystalline nodal topological phase happens close to 
the \(\alpha_Rp_F=b\) line.

Along the $b=0$ line, the nodes on the $\Gamma$
pocket in the \(\mu = z\) triplet state are lifted due to the admixture
with the sub-leading s-wave \(\mu = 0\) state generated by Ising SOC, 
resulting in an anisotropic gap.
This is shown in Fig. \ref{Gaps}(c), which presents the evolution
of $\Delta_{\Gamma1}\left(\mathbf{p}\right)$ along the $b=0$ axis
for increasing $\alpha_{R}$.
Note that to generate singlet-triplet mixing, we must include a small
difference between the inner and outer DOS in the plots of Fig. \ref{Gaps}; the magnitude of this mixing increases with $\alpha_R$.
Thus, although the superconducting gap on $\Gamma$ in the triplet regime remains strongly modulated as a function of angle for modest $\alpha_R$ (see Fig. \ref{Gaps}(c)), the superconducting phase in this region is nonetheless fully gapped.

For large values of $\alpha_{R}$ (of the order of the Ising SOC), the dominant instability shifts from being in the $A_1$ irrep of $C_{3v}$ (previously the $A_1'$ irrep of the original $D_{3h}$ group) to the 2-dimensional $E$ irrep (previously the $E''$ irrep of the original $D_{3h}$ group). As a result, a new chiral $p\pm ip$ superconducting state emerges in the triplet regime at $b=0$. %{\color{cyan}(see Appendix \ref{B} for a more detailed discussion of the critical temperatures)}. 
The chiral phase, shown in Fig. \ref{Gaps}(d), occurs
%because the gap formally transforms as
%a two-dimensional irreducible representation \(E\) of the relevant $C_{3v}$ point group.
%{\color{blue}as a result of the degeneracy of the (x) and (y) solutions being resolved at the fourth order of the free energy,}
because in this 2-dimensional irrep the free energy is minimized by a spontaneous breaking of  time reversal symmetry, as we discuss in Appendix \ref{C}. This is in agreement with the general result in \cite{Samokhin15,Schmalian17}.  
As we show in Appendix \ref{C}, this results in a gapped chiral topological SC with
gapless chiral edge modes resulting in a thermal
Hall conductance $\kappa_{xy}=\pm6\left(\pi^{2}k_{B}^{2}/3h\right)T$
\cite{ReadGreen}. This topological SC phase survives for sufficiently small in-plane magnetic fields, %for some range of $b$,
but our approach is insufficient to quantitatively obtain
the phase boundary (see blue dashed line in Fig. \ref{PhaseFig}(b)).

\subsection{Broken Momentum Inversion Symmetry: Bogolyubov Fermi Surfaces and Finite-Momentum Pairing}\label{SecIVC}

In the presence of both $\alpha_R$ and $b$, the Fermi surfaces are no longer inversion-symmetric under \(\mathbf{p}\rightarrow-\mathbf{p}\). The key quantity that measures the degree of symmetry breaking is
\begin{equation}
\xi_{A\eta\tau}(\mathbf{p})= \frac{\xi_{\eta\tau}(\mathbf{p})-\xi_{-\eta\tau}(-\mathbf{p})}{2}
\end{equation}
previously defined in Eq. (\ref{Eq:Xi}).  Here we discuss two important consequences of this inversion symmetry breaking for the crystalline topological SC state.  
First, within the crystalline topological SC state, %{\color{blue}that occurs when \(b>\alpha_Rp_F\)},
breaking inversion symmetry moves the two nodes on the same Fermi surface at \(\Gamma\) in opposite directions away from the Fermi level. This follows directly from Eq. (\ref{Eq:BdGspec}) as the nodes move by an energy \(\xi_{A\Gamma\tau}(\mathbf{p}_{node})\). 
As we showed in Fig. \ref{GapFig}(c), this results in the nodes `inflating' into Bogolyubov Fermi surfaces(\emph{cf}. \cite{AgterbergBFS17,AgterbergBFS18,FuBFS18,Yanase19}).  These Fermi surfaces are protected by mirror symmetry, due to the topological stability of the band crossings in the BdG spectrum, as we show in the next section.

\begin{figure}
\centering \includegraphics[width=0.99\columnwidth]{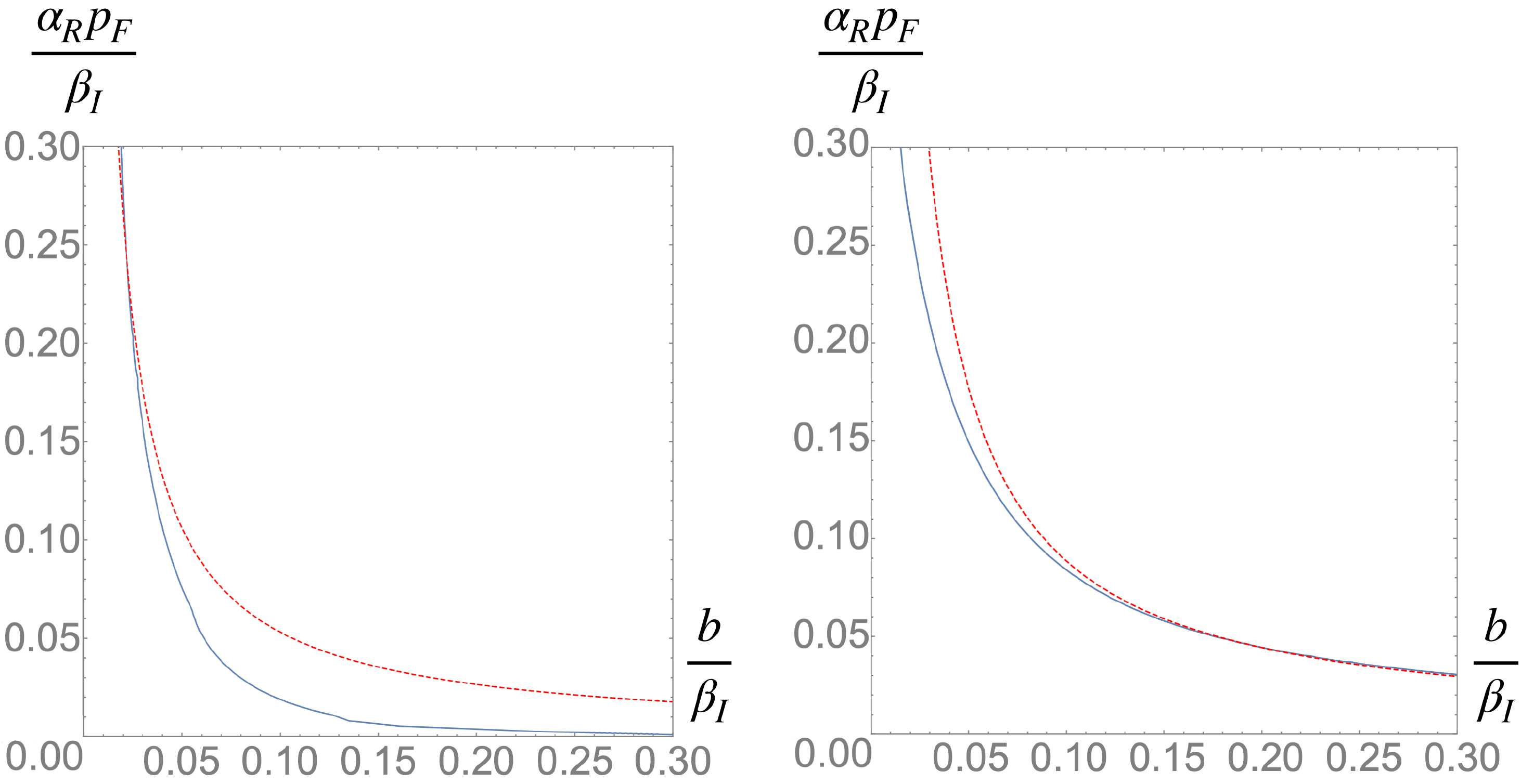} \\
\caption{Critical lines above which uniform SC becomes unstable for the singlet (left) and triplet (right) phase diagrams shown in Fig. \ref{PhaseFig}. Blue line is the numerical solution from the full gap equation, while the red dashed line is given by the approximation \(\alpha_Rp_F b=\frac{\beta_I T_{c0}}{1.13}\).  In addition to the parameters used in Fig. \ref{Gaps}, we took \(T_c=0.01\beta_I\) and \(\Lambda=25\beta_I\), roughly corresponding to the observed values \cite{MakNat16,Hunt,Khodas18}.}
\label{StabCurves} 
\end{figure}

Second, breaking momentum inversion symmetry cuts off the Cooper logarithm in the particle-particle bubble in Eq. (\ref{PiXiA0}). As a result, the pairing interaction must be larger than a certain threshold, proportional to how much inversion symmetry is broken, for a uniform SC state to onset.  To show this explicitly, we evaluate the particle-particle bubble in Eq. (\ref{Pi}) in the absence of inversion symmetry.  
Assuming that \(\xi_{A\eta\tau}\) is a function only of the direction \(\theta\) around the Fermi surface,
in the limit of \(\Lambda\gg\xi_{A\eta\tau}\) we find
\begin{equation}
\frac{\Pi_{\eta\tau}(\theta)}{N_{\eta\tau}}=-\ln\frac{1.13\Lambda}{T_c}+\text{Re}\left[\psi\left(\frac{1}{2}+\frac{i\xi_{A\eta\tau}(\theta)}{2\pi T_c}\right)-\psi\left(\frac{1}{2}\right)\right]
\end{equation}
where \(\psi\) is the digamma function.  As a result, at zero temperature
\begin{equation}\label{Eq:PiT0}
\Pi_{\eta\tau}(\theta)=-N_{\eta\tau}\ln\frac{\Lambda}{|\xi_{A\eta\tau}(\theta)|}
\end{equation}
i.e. the infrared logarithmic divergence originally present is cutoff by \(|\xi_{A\eta\tau}(\theta)|\). This means that there is a critical value of the parameter \(\xi^{c}_{A\eta\tau}\) beyond which uniform SC is no longer stable. The resulting critical lines are shown in Fig. \ref{PhaseFig} and for a larger range in Fig. \ref{StabCurves} for both singlet and triplet instabilities. Note that because this is a multi-band superconductor, the critical line has a (weak) dependence on the cutoff \(\Lambda\).

The general shape of the critical lines can be understood from a simple approximation, noting that
\begin{equation}
\xi_{A\eta\tau}(\mathbf{p})=\frac{\tau}{2}\left(\delta_\eta(\mathbf{p})-\delta_{-\eta}(-\mathbf{p})\right),
\end{equation}
where \(\delta_\eta\) are functions of \(\alpha_Rp_F\) and \(b\) as given in (\ref{delta}). For \(\alpha_Rp_F,b\ll\beta_I\), 
\begin{equation}
\xi_{A\eta\tau}(\mathbf{p})\approx \tau \frac{\alpha_Rp_F b}{\beta_{I}}\sin(\theta-\vartheta)
\end{equation}
where \(\beta_{I}\) is the Ising SOC on the relevant pocket and $\vartheta$ is the direction of the magnetic field. From (\ref{Eq:PiT0}), we can estimate that the characteristic scale of \(\xi^{c}_{A\eta\tau}\), properly averaged, is of the order \(T_{c0}/1.13\), where $T_{c0}$ is the solution of the gap equation when  \(\xi^{c}_{a\eta\tau}=0\), see Eq. (\ref{PiXiA0}). The critical curve is thus roughly given by
\begin{equation}
\alpha_Rp_F b\sim\frac{\beta_I T_{c0}}{1.13}
\end{equation}
As shown in Fig. \ref{StabCurves}, this approximation reasonably captures the exact result for the critical line.

For inversion symmetry breaking exceeding the critical value, where the uniform SC state is no longer stable, superconductivity with finite center-of-mass momentum \(\mathbf{p}_\mathrm{{shift}}\neq0\), i.e. a so-called FFLO phase \cite{AgterbergFFLO, pipStarykh,Liu17}, is still possible. Note that the momentum shift is not necessarily equal for each Fermi surface, so more generally there are four parameters \(\mathbf{p}_{\mathrm{shift},\eta\tau}\).
Depending on whether \(\mathbf{p}_{\mathrm{shift},\eta\tau}=\mathbf{p}_{\mathrm{shift},\eta-\tau}\) or \(\mathbf{p}_{\mathrm{shift},\eta\tau}=-\mathbf{p}_{\mathrm{shift},\eta-\tau}\), the FFLO phase is classified as helical and stripe, respectively, and may even compete with the uniform SC phase below the threshold curve in the $(\alpha_R p_F, b)$ plane \cite{AgterbergFFLO}. Ultimately, the  four parameters \(\mathbf{p}_{\mathrm{shift},\eta\tau}\) must be obtained by minimization of the free energy, which is a computationally involved task beyond the scope of our work. It is interesting to note, however, that by matching  \(\mathbf{p}_\mathrm{{shift}}\) with the geometric shift of the center of the corresponding Fermi surface, 
the nodes of the superconducting ground state move back to the Fermi level since the shift compensates for the finite \(\xi_{A\eta\tau}\) (see Eqs. (\ref{BdGFFLO})-(\ref{Eq:XiFFLO}) below), as shown in Fig. \ref{GapFig}(d).  
Because this configuration maximizes the gap around the Fermi surface, it is expected to maximize the condensation energy. In any case, as we show in the next section, the finite momentum pairing does not affect the topological properties of the SC phase.

\section{Crystalline gapless topological superconductivity}\label{V}

Having established  
the existence of a nodal SC phase for large magnetic fields in the
phase diagrams of Fig. \ref{PhaseFig}, we now discuss its topological
properties. As discussed in Refs. \cite{RyuPRB11,SatoPRB11,RyuNJP13,Schnyder14,Sato14,Ryu16rev},
two-dimensional gapless topological phases are stable only in the presence
of certain symmetries, which guarantee stability of both the bulk
nodes and of the corresponding edge modes. When $\alpha_{R}=0$,
the SC state has both particle-hole symmetry \(\mathcal{C}\) and an anti-unitary time-reversal-like
symmetry $\tilde{\mac{T}}=i\sigma^{x}\mathcal{K}$ ($\mathcal{K}$
is complex conjugation, and $\sigma^{x}$ acts on the spin index),
which is a composition of time-reversal symmetry and a reflection
with respect to the $xy$ plane. $\tilde{\mac{T}}$ reverses the in-plane
momentum and the $z$ component of the spin, satisfying $\tilde{\mac{T}}^{2}=1$.
This time-reversal-like symmetry places the system into symmetry class
BDI \cite{AltlandZirnbauer,RyuNJP10,Ludwig15} and protects the 12 nodes of
the superconducting gap on the two $\Gamma$ pockets along the $\Gamma$-$M$
lines, ensuring that the boundary flat bands cannot be gapped \cite{RyuNJP13,Agterberg18}.
However, a finite Rashba SOC breaks the $\tilde{\mac{T}}$ symmetry, putting the model in symmetry class D.   
For generic in-plane field directions, this results in a fully gapped, topologically trivial SC phase with no protected
zero-energy boundary states.

The notable exception is when ${\bf B}$ is parallel to one of the
$\Gamma$-$K$ directions: in this case, the system has a mirror symmetry
associated with reflection about the plane perpendicular to ${\bf B}$.  
For example, when $\mathbf{B}$ is parallel to the $x$ axis, the
mirror symmetry corresponds to a reflection with respect to the $yz$
plane perpendicular to ${\bf B}$, which also flips the $y$ and $z$
components of spin: $\mac{M}_{x}=i\sigma^{x}R_{yz}$, where the reflection $R_{yz}$
corresponds to $(x,y,z)\rightarrow(-x,y,z)$ and, as above, $\sigma^{x}$
acts on the physical spin indices.  
In general the superconducting gap can be either even or odd under this mirror symmetry. 
We show below that for $b > \alpha_R p_F$, where the gap is mirror-odd, 
the reflection symmetry anti-commutes with particle-hole symmetry.   
In this case, there is a \(\mathbb{Z}\)-valued topological invariant that diagnoses the topological superconducting phase \cite{Schnyder14}. 
  %which we show is non-trivial in the reflection-symmetric superconductor with $b > \alpha_R p_F$. %which we discuss in more detail in the following section (\(M\) indicates that it is protected by the mirror symmetry). 
This invariant is characterized by non-vanishing quantized winding number along a contour encircling each node -- or if the nodes are not at the Fermi energy \cite{RyuNJP10}, a contour encircling each Bogolyubov Fermi surface \footnote{In the case of multiple bands, each node changes the \(M\mathbb{Z}\) invariant by \(\pm 1\), depending on which bands cross}.
Thus mirror reflection protects the
four nodes in the reflection plane provided that $b > \alpha_R p_F$ (see Fig. \ref{GapFig}(c))\cite{Schnyder14,Sato14,Sato17}.  When $b = \alpha_R p_F$, pairs of nodes touch and there is a topological phase transition into a nodeless phase at $b < \alpha_R p_F$.

We emphasize
that the topological nature of this  $\alpha_{R}\neq0$ SC state is
qualitatively different than that of the $\alpha_{R}=0$ SC state, as in
the former case the symmetry that protects the SC state is not time-reversal-like,
but a mirror symmetry \textendash{} hence the denomination
crystalline gapless topological SC \cite{Schnyder14,Sato14}.

\subsection{Mirror Symmetry and Topologically Protected Nodes}

To see how the mirror reflection symmetry protects the nodes in the reflection plane, we will follow the approach of Ref. \cite{Schnyder14}, who established that an analysis of the symmetry-allowed mass terms in an effective low energy theory can reveal whether a non-gapped superconductor is topologically non-trivial.   
%To examine the topological crystalline superconducting phase, we take the magnetic field to be along the $\hat{\mathbf{x}}$ direction, in which case the mirror symmetry 
%\(\mathcal{M}_{x}\) is reflection in the $y-z$ plane perpendicular to the superconducting layer. 
\(\mathcal{M}_{x}\) acts on the non-interacting Hamiltonian (\ref{Eq:H0}) as
\begin{equation} \mathcal{M}^{-1}_{x}H(\mathbf{p}) \mathcal{M}_{x} = H(\bar{\mathbf{p}})\end{equation}
 where \(\bar{\mathbf{p}}=(-p_x,p_y)\). Since this reflection also reverses the $y$ and $z$ components of the spin, in the spin basis \(\mathcal{M}_{x}\) acts as \(i\sigma^x\), while in the SOC basis (\ref{Cbasis}) it is momentum dependent, \(\mathcal{M}_{x}(\mathbf{p})=U^\alpha_{\eta\tau}(\mathbf{p})i\sigma^x_{\alpha\beta} U^{\beta*}_{\eta\tau'}(\bar{\mathbf{p}})= i\tau e^{-i\phi(\mathbf{p})}\delta_{\tau\tau'}\), where $\phi(\mathbf{p})$ is given in Eq. (\ref{delta}).

The action of mirror symmetry can be extended to the BdG spinors to give 
\begin{eqnarray} \label{Eq:FullMirror}
\tilde{\mathcal{M}}_{x}(\mathbf{p})=\left(\begin{array}{cc}
\mathcal{M}_{x}(\mathbf{p}) & 0\\
0 & - \mathcal{M}^\dagger_{x}(-\mathbf{p}) 
\end{array}\right  )\nonumber \\
=\left(\begin{array}{cc}
e^{i\phi(\mathbf{p})} & 0\\
0 & - e^{-i\phi(-\mathbf{p})}
\end{array}\right)
\end{eqnarray}
On the BdG Hamiltonian Eq. (\ref{BdG}),
%\begin{equation}\label{BdG}
%\mathcal{H}_{\eta\tau}(\mathbf{p})=\left(\begin{array}{cc}
%\xi_{\eta\tau}(\mathbf{p}) & \Delta_{\eta\tau}(\mathbf{p}) \\
%\Delta_{\eta\tau}^*(\mathbf{p}) & -\xi_{-\eta\tau}(\mathbf{-p})
%\end{array}\right) ,
%\end{equation}
the mirror symmetry thus acts according to
\be \label{BdGMirror}
\tilde{\mathcal{M}}_{x}^{-1} \mathcal{H}_{\eta\tau}(\mathbf{p}) \tilde{\mathcal{M}}_{x} =\mathcal{H}_{\eta\tau}(\bar{\mathbf{p}}) \ .
\ee
Here the relative sign between the two non-vanishing components of $\tilde{M}$ is fixed by whether the gap function is even or odd under the mirror symmetry.  For $b > \alpha_R p_F$, where the SC gap that gives the highest \(T_c\) in Eq. (\ref{Eq:Gap})  is odd under $p_x \rightarrow - p_x$, the appropriate choice is minus \footnote{There is also a solution of the gap equation that is even under $p_x \rightarrow - p_x$, namely the \((x)\) channel, which therefore decouples from the odd solution by symmetry, but which can be shown to always have lower \(T_c\).}. 
%(the rest of the \((\mu)\) terms in Eq.'s (\ref{Eq:GapFuns},\ref{GapFnc}) with \(\mu=0,y,z\) can directly be shown to be odd for $b > \alpha_R p_F$; for $b < \alpha_R p_F$, intechange even and odd)

Following the methods of Ref. \cite{Schnyder14}, we will show that in the low-energy theory obtained by linearizing the model near the nodes, there are no symmetry-allowed mass terms; this is equivalent to showing that the topological winding number is non-trivial (which we have also verified in our model by directly computing the Berry connection, though we will not present the calculation here).  However, we do find symmetry-allowed terms that shift the nodes away from the Fermi energy, leading to protected Bogolyubov Fermi surfaces.  

We begin by linearizing the Hamiltonian in the region $b > \alpha_R p_F$ around the pair of nodes at \(p_x=0\). This gives a $4 \times 4$ low-energy effective Hamiltonian, with a new index  $L, R$ to keep track of the two nodes.  We define \(\tau^\mu\) to be the Pauli matrices acting on the $L,R$ indices, while $\varsigma^\mu$ are Pauli matrices acting on the 2 indices of the BdG spinors (i.e. on the particle-hole indices). 
In this basis the particle-hole symmetry, which interchanges the two nodes, acts via \(\mathcal{C}=\varsigma^x\otimes\tau^x\mathcal{K}\). 
%which reduces to \(\varsigma^z\) for \(b>\alpha_Rp_{F}\), when our system is in the nodal phase.  
Note that Eq. (\ref{Eq:FullMirror}) implies that the action of $\mathcal{M}_{x}$ on the mirror plane changes discontinuously at $b = \alpha_R p_F$: for $b<\alpha_Rp_{F}$, $\mathcal{M}_{x}$ is proportional to the identity matrix times \(\sgn p_y\), while for $b>\alpha_Rp_{F}$ it has the form
$\tilde{\mathcal{M}}_{x}(\mathbf{p})=\varsigma^z$.
Since the mirror symmetry acts in the same way near both nodes, in our linearized theory it acts via 
\be
\tilde{\mathcal{M}}_{x}=\varsigma^z\otimes\tau^0 \ . 
\ee
Since the nodes are in the mirror plane, this leading-order approximation is sufficient to determine which terms can open a gap, 

We now consider all leading terms of the generic form \(h(\delta p_y,p_x)\varsigma^\mu\otimes\tau^\nu\) allowed by symmetry, with $\delta p_y = p_y - p_y^{(\text{node})}$. Note that we do not wish to allow terms that couple the two nodes, as these break translational symmetry; thus we require $\nu = 0$ or $z$.   Recall that the symmetries are 
\ba
\mathcal{C}^{-1}\mathcal{H}(\delta p_y,p_x)\mathcal{C}=-\mathcal{H}(-\delta p_y,-p_x) \n
\tilde{\mathcal{M}}_{x}^{-1}\mathcal{H}(\delta p_y,p_x)\tilde{\mathcal{M}}_{x}=\mathcal{H}(\delta p_y,-p_x) \ \ .
\ea 
Thus \(h(-\delta p_y,-p_x) =   \pm h(\delta p_y,p_x)\) when $\mathcal{C}^{-1} \varsigma^\mu\otimes\tau^\nu\mathcal{C} = \mp  \varsigma^\mu\otimes\tau^\nu$.  Similarly  \(h(\delta p_y,-p_x) =  \pm h(\delta p_y,p_x)\) when $\tilde{\mathcal{M}}_{x}^{-1} \varsigma^\mu\otimes\tau^\nu\tilde{\mathcal{M}}_{x}= \pm  \varsigma^\mu\otimes\tau^\nu$.  To gap out the nodes we must have \(h(0,0)\neq 0\); thus we need plus signs in both cases.  Hence \(\varsigma^\mu\otimes\tau^\nu\) anti-commutes with $\varsigma^x\otimes\tau^x\mathcal{K}$ and commutes with $\tilde{\mathcal{M}}_{x}$.

This restricts the linearized Hamiltonian for \(b>\alpha_Rp_{F}\) to the form
\begin{equation}\label{LinH}
\mathcal{H}=\delta p_y\varsigma^z\otimes\tau^z+p_x\varsigma^x\otimes\tau^0+m\varsigma^z\otimes\tau^0+\xi_A \varsigma^0\otimes\tau^z
\end{equation} where \(m\) and \(\xi_A\) are constants (for simplicity, we take the Fermi velocity to be \(1\), so all parameters have the same units). The \(m\) term, which plays the role of a chemical potential shift at each node, does not lift the nodes; rather it shifts them in opposite directions along the $p_y$ axis, from \(p_y=\pm p_{F}\) to  \(p_y=\pm (p_{F}+m)\).
The \(\xi_A\varsigma^0\otimes\tau^z\) term, on the other hand, shifts the nodes in opposite directions in energy by an amount \(\xi_A\), inflating the nodes into small Bogolyubov Fermi surfaces (see also \cite{AgterbergBFS17,AgterbergBFS18,FuBFS18,Yanase19}).  %Comparing with Eq. (\ref{BdG}) and Eq. (\ref{Eq:BdGspec}), a short computation shows \begin{equation}\label{Eq:BdGspec}
%E_{\eta\tau}(\mathbf{p})=\xi_{A\eta\tau}(\mathbf{p})+\sqrt{\xi_{S\eta\tau}(\mathbf{p})^2+|\Delta_{\eta\tau}(\mathbf{p})|^2}
%\end{equation}
%By examining the eigenvalues of the BdG Hamiltonian given in
Comparing with the BdG spectrum Eq. (\ref{Eq:BdGspec}), we find that \(\xi_A\) is precisely the value of \(\xi_{A\Gamma\tau}\) given in Eq. (\ref{Eq:Xi}) when \(\xi_{S\Gamma\tau}({\bf p}) =0\) in Eq. (\ref{Eq:XiS}). Since a constant shift in energy cannot change the Berry connection, the winding numbers are unaffected and remain non-trivial \cite{Ryu16rev,Sato14}, as can be verified by direct computation. %This is in accord with the fact that only the ``minimal" dimension of the Fermi surface in the gapless phase determines the topology of the phase cite{Ryu16rev}, which follows from the K theoretic classification in which only the dimension of a hypersphere enclosing the topological defect/Fermi surface is relevant \cite{Sato14}.
Note that the Bogolyubov Fermi surfaces are  topologically protected only in a fragile sense \cite{Chapman19}, as they can be removed by mixing with additional bands, 
similar to what has been observed theoretically in 1D crystalline topological insulators \cite{Bernevig11,Carmine18}. %As in those systems, we expect some experimental signatures to remain even when the Bogolyubov Fermi surfaces are removed.  

%To see that \(\xi_A\varsigma^0\otimes\tau^z\)  describes a normal state Fermi surface that is shifted 
%by a momentum of \(\xi_A\) relative to the Brillouin zone center, note that at \(p_x=0\), the FS at both nodes is determined by \(\delta p_y +\xi_A=0\), which describes the $p_x =0$ projection of a BdG Fermi surface for non-zero $\xi_A$. %%ds: I couldn't figure out what was being said here, so I replaced it with the text used in the previous section

The term \(\xi_A\varsigma^0\otimes\tau^z\) describes a state with broken momentum inversion symmetry, in which the normal state Fermi surface is shifted by a momentum of \(\xi_A\) relative to the Brillouin zone center. To see this, note that at \(p_x=0\), the normal state Fermi surface at both nodes is determined by \(\delta p_y +\xi_A=0\). This means that Cooper pairs cannot both lie on the Fermi surface.
As noted in Sec. \ref{SecIVC}, for sufficiently large inversion symmetry breaking, uniform pairing becomes unstable, but an FFLO-type finite-momentum pairing  between electrons \emph{on} the Fermi surface remains a possibility \cite{AgterbergFFLO}. In this case the Nambu spinors have to be redefined as \((c_{\mathbf{p}+p_\mathrm{{shift}}\hat{\mathbf{y}}},c^\dagger_{-\mathbf{p}+p_\mathrm{{shift}}\hat{\mathbf{y}}})\), where $2p_\mathrm{{shift}}$ is the total momentum of the pair.
This transforms the BdG Hamiltonian in Eq. (\ref{BdG}) into
\begin{equation}\label{BdGFFLO}
\mathcal{H}_{\eta\tau}(\mathbf{p})=\left(\begin{array}{cc}
\xi_{\eta\tau}(\mathbf{p+p}_\mathrm{{shift}}) & \Delta_{\eta\tau}(\mathbf{p}) \\
\Delta_{\eta\tau}^*(\mathbf{p}) & -\xi_{-\eta\tau}(\mathbf{-p+p}_\mathrm{{shift}})
\end{array}\right).
\end{equation}
with the spectrum now given by
\begin{equation}\label{BdGSpecFFLO}
E_{\eta\tau}(\mathbf{p})=\xi_{A\eta\tau}(\mathbf{p,p}_\mathrm{{shift}})+\sqrt{\xi_{S\eta\tau}(\mathbf{p,p}_\mathrm{{shift}})^2+|\Delta_{\eta\tau}(\mathbf{p})|^2}
\end{equation}
with
\begin{eqnarray}\label{Eq:XiFFLO}
\xi_{S\eta\tau}(\mathbf{p,p}_\mathrm{{shift}})=\frac{\xi_{\eta\tau}(\mathbf{p+p}_\mathrm{{shift}})+\xi_{-\eta\tau}(-\mathbf{p+p}_\mathrm{{shift}})}{2}\nonumber\\
\xi_{A\eta\tau}(\mathbf{p,p}_\mathrm{{shift}})=\frac{\xi_{\eta\tau}(\mathbf{p+p}_\mathrm{{shift}})-\xi_{-\eta\tau}(-\mathbf{p+p}_\mathrm{{shift}})}{2}\nonumber\\
\end{eqnarray}
Linearizing the transformed BdG Hamiltonian around the nodes (where $\xi_{S\eta\tau}(\mathbf{p,p}_\mathrm{{shift}}) =0$)
yields a new term \(-p_\mathrm{{shift}}\varsigma^0\otimes\tau^z\) in Eq. (\ref{LinH}).  Picking \(p_\mathrm{{shift}}=\xi_A\), this term cancels the energy shift of the nodes, bringing them back to the Fermi level.
%This amounts to setting \(\xi_{A\eta\tau}(\mathbf{p,p}_\mathrm{{shift}})\) to be zero simultaneously with \(\xi_{S\eta\tau}(\mathbf{p,p}_\mathrm{{shift}})\).

%{\fb To add here: a note about parameters used in Fig. 4, and a discussion of why we show small Rashba so as not to have to account for FS shifts.}

\begin{figure}
\centering \includegraphics[width=0.98\columnwidth]{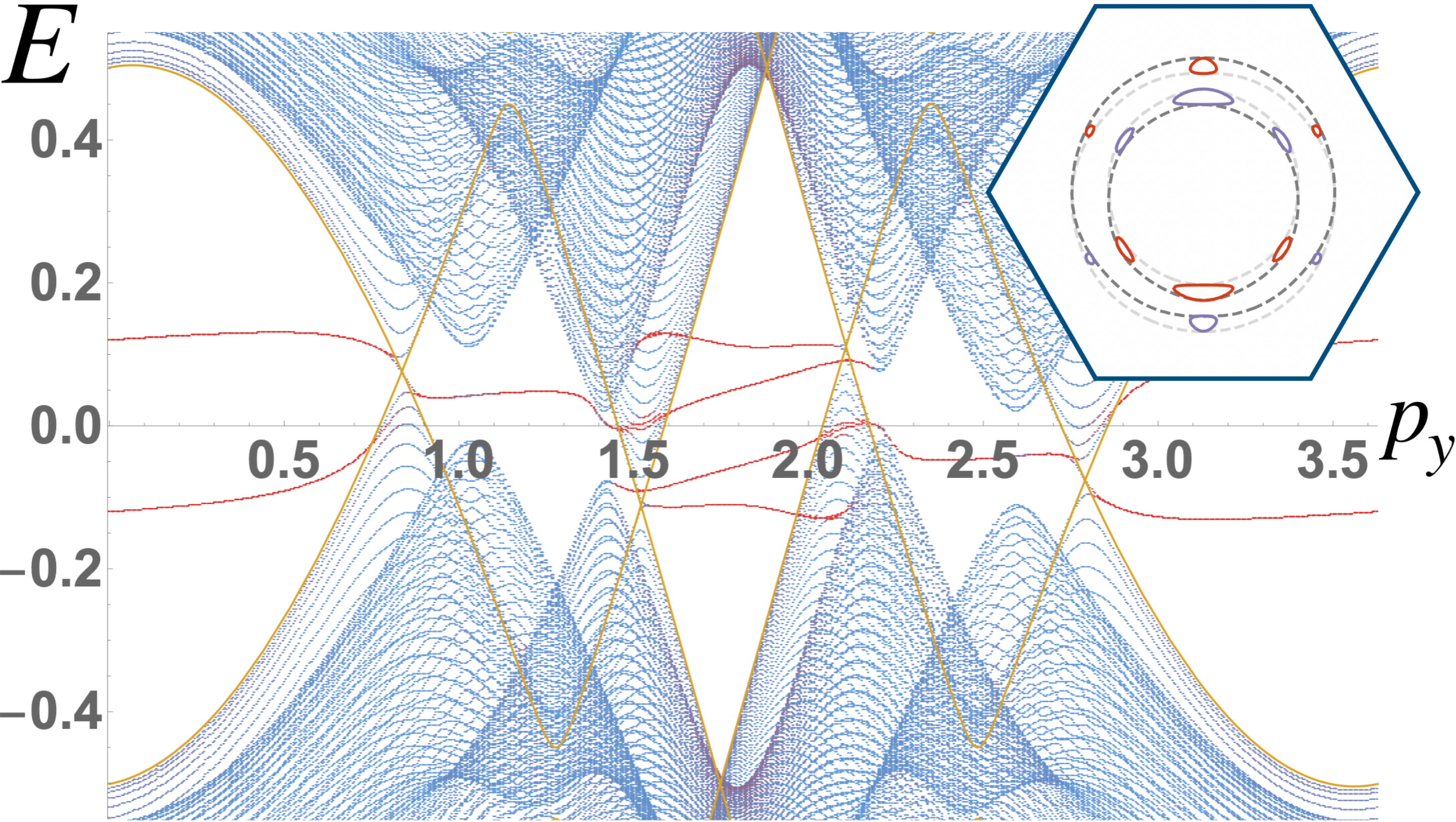}
\caption{Excitation spectrum in the topological crystalline
SC phase on a $150\times\infty$ unit cell strip with ${\bf B}=B\hat{{\bf x}}$.
Blue indicates delocalized bulk eigenstates, red indicates eigenstates concentrated
near the boundaries, and yellow shows a cut of the bulk BdG spectrum at $p_x=0$. The inset illustrates the Bogolyubov Fermi surfaces
and the original Fermi surfaces (dashed black lines) and its inverse image under \(\mathbf{p}\rightarrow-\mathbf{p}\) (dashed gray lines). %Only surfaces crossing the \(p_y\) axis are protected.
For details of the tight-binding model, see Appendix \ref{TightBinding}.}
\label{CylFig} % \end{center}
\end{figure}

\subsection{Boundary modes}

By the bulk-boundary correspondence \cite{Schnyder14,Sato14}, there are edge bands terminating at the nodes.  However, unlike other nodal topological superconductors with Majorana flat band edge modes \cite{LawMay16,RyuPRB11,SatoPRB11,Sato18}, the edge modes in the crystalline topological phase under consideration are not in general flat and not necessarily at zero energy.   The edge bands can be studied following the methods of  \cite{Witten15,Wu17}; we find that for open boundary conditions in $x$ and \(p_y\) close to the node, the edge mode has energy \(\xi_A\).   When \(\xi_A\) vanishes at the node (e.g. if a stripe FFLO phase is realized in the bulk) this means that the boundary modes cross zero energy--  but they are not flat in general as \(\xi_A\) does not have to vanish for all \(p_y\). Similar edge states have been studied in 3D crystalline topological insulators, where they are referred to as drumhead states \cite{Zahid16,Schnyder16,WangRyu18}.

An alternative way to understand the edge modes, is to view them as topological boundary modes of  the family of 1D Hamiltonians \(\mathcal{H}_{p_y}(p_x)=\mathcal{H}(p_x,p_y)\) at fixed \(p_y\), which are in topological class A and respect mirror symmetry. Since mirror symmetry is equivalent to inversion in 1D, these are the same systems as studied in \cite{Bernevig11,Carmine18}. As \(p_y\) crosses the node, \(\mathcal{H}_{p_y}(p_x)\) undergoes a topological phase transition from trivial to non-trivial. The 1D topological invariant is the mirror index \(M\mathbb{Z}\) defined as follows \cite{Bernevig11,Ryu13}.  Since \(\mathcal{H}_{p_y}(p_x)\) commutes with \(\mathcal{M}_x\) at \(p_x=0\) and \(p_x=\pi\) (the boundary of the 1D Brillouin zone), at those points it can be decomposed into two blocks on which \(\mathcal{M}_x=\pm1\) respectively: \(\mathcal{H}^\pm_{p_y}(0)\) and \(\mathcal{H}^\pm_{p_y}(\pi)\) . If \(n_+^{(0)}\) is the number of occupied states of \(\mathcal{H}^+_{p_y}(0)\) and \(n_+^{(\pi)}\) is that of \(\mathcal{H}^+_{p_y}(\pi)\), the mirror index is simply \(N_{M\mathbb{Z}}=|n_+^{(0)}-n_+^{(\pi)}|\). There are then \(2N_{M\mathbb{Z}}\) edge states, with each state even under reflection being degenerate with an edge state that is odd under reflection. Taken together these edge states form the band of edge modes of the 2D system. At a node (which is at \(p_x=0\)), \(\mathcal{H}^+_{p_y}(0)\) crosses with a state in \(\mathcal{H}^-_{p_y}(0)\), which changes \(n_+^{(0)}\) by one and the number of edge states by two. The edge mode thus splits into two bulk modes which cross at the node.

To study the boundary modes of our model in the uniform superconducting state, in Appendix \ref{TightBinding} we describe a tight-binding model that captures the key features of our gapless topological superconducting phase.  The results are displayed in 
 Fig. \ref{CylFig}, which shows the Bogolyubov-de
Gennes (BdG) spectrum on a $150\times\infty$ unit cell strip with
open zig-zag edges parallel to the $\hat{y}$ direction, and ${\bf B}=B\hat{x}$, in the uniform superconducting state.
Each state $\psi_{k}$ is colored according to the inverse participation
ratio $\sum_{y}|\psi_{k}(y)|^{4}$, such that the boundary modes
are red, and bulk modes are blue.
A cut containing the nodes along $p_x=0$ of the bulk BdG spectrum is shown in yellow. 
%Unlike in other nodal topological superconductors \cite{LawMay16,RyuPRB11,SatoPRB11,Sato18}, the edge modes are not pinned to zero energy, and are not in general flat 

It is worth noting that in actual materials, the existence of both bulk nodes and corresponding boundary states is guaranteed only
if the relevant mirror reflection is an exact symmetry. As such, these
may be sensitive to orientational defects in the crystal.

\section{Concluding remarks}\label{VI}

Our microscopic interacting model for
NbSe$_{2}$ predicts multiple possible exotic superconducting phases
in this material, tuned by the Rashba SOC $\alpha_{R}$ and the in-plane
magnetic field $B$. Two different primary SC instabilities (i.e. that would take place when Ising SOC is zero) can be
driven by purely electronic interactions: a singlet extended $s$-wave
and a triplet $f$-wave instability.   The phase diagrams for both are qualitatively similar, with a fully gapped 
superconductor for $\alpha_R > b$, and a  crystalline gapless topological SC state for large b and small $\alpha_R$.  Interestingly, the topological
properties of the latter phase depend crucially on the $\mathbf{B}$
field being aligned along one of the $\Gamma$-$K$ directions.
In addition, the triplet instability supports
a chiral topological SC state for small $b$ and large $\alpha_{R}$.

Although direct experimental detection of these topological SC states
may be technically challenging, their indirect
experimental manifestations should be accessible. For instance, because
the chiral SC state transforms as a two-dimensional irreducible representation
of the trigonal space group, it should be strongly affected by strain,
with $T_{c}$ splitting into $2$ separate transitions under externally
applied uniaxial strain \cite{Hicks14}. As for the crystalline topological
SC state, its extreme sensitivity to the field direction is expected
to promote strongly anisotropic properties. Specifically, since the nature of the SC state changes as a function of the B direction, one expects pronounced six-fold anisotropies in the upper critical field and in the critical current. Recent experiments have identified six-fold and two-fold anisotropies in the magneto-resistance and in the critical field \cite{Pribiag20, Lortz20}. Whether these results can be attributed to the crystalline nodal topological superconducting state discussed here requires further investigation. Finally, the presence of Bogolyubov Fermi surfaces should also be manifested in several experimental observables that are sensitive to the existence of a  finite DOS at zero energy \cite{Menke19}.

\begin{acknowledgments}
We thank T. Birol, A. Chubukov, V. Pribiag, and K. Wang for fruitful
discussions. This work was supported primarily by the National Science Foundation (NSF) 
Materials Research Science and Engineering Center at the University of Minnesota 
under Award No. DMR-1420013, via an iSuperSeed Award (DS, FJB, and RMF). FJB is grateful
for the financial support of the Sloan Foundation FG-2015- 65927.
JK was supported by the National High Magnetic Field Laboratory through
NSF Grant No. DMR- 1157490 and the State of Florida.
\end{acknowledgments}

\bibliography{BibTeXforSCinTMDs}

\appendix

\section{Ginzburg-Landau Free Energy}\label{A}

Here we write down the Ginzburg-Landau free energy in the presence of SOC and magnetic field. %needed to derive the gap equation (\ref{Eq:Gap}), including the particle-particle susceptibility (\ref{Pi}), assuming well-separated inner and outer Fermi surfaces or sufficiently weak pairing
We later use it in Appendix \ref{C} to analyze the chiral phase that emerges in the regime of dominant triplet interactions in the absence of magnetic field and large values of Rashba SOC.

We start with the Bogolyubov-Gor'kov Hamiltonian obtained after doing a Hubbard-Stratonovich transformation:
\begin{equation}
H=\frac{1}{2}\sum_{\mathbf{p}\eta\tau}\Psi_{\mathbf{p}\eta\tau}^\dagger\mathcal{H}_{\eta\tau}(\mathbf{p})\Psi_{\mathbf{p}\eta\tau}+\frac{1}{2}\sum_{\mathbf{p}\eta\tau} \xi_{\eta\tau}(\mathbf{p})+\mathcal{H}_{0}\left(\Delta^{2}\right)
\end{equation}
where
\begin{equation}\label{Eq:mathcalH0}
\mathcal{H}_{0}\left(\Delta^{2}\right)=-\frac{1}{4}\sum_{\substack{\mathbf{p}\eta\tau \\\mathbf{k}\eta'\tau'} }\Delta_{\eta\tau}^*(\mathbf{p})\left(\tilde{V}^{-1}(\mathbf{p};\mathbf{k})\right)^{\eta'\tau'}_{\eta\tau}\Delta_{\eta'\tau'}(\mathbf{k})
\end{equation}
and where we use the Nambu-Gor'kov representation \(\Psi_{\eta\tau}(\mathbf{p})=(c_{\eta,\mathbf{p}\tau},c^\dagger_{-\eta,\mathbf{-p}\tau})^T\) and the BdG Hamiltonian Eq. (\ref{BdG}). Recall that the BdG spectrum has two branches, \(E_{\eta\tau}(\mathbf{p})\) and, by particle-hole symmetry, \(-E_{-\eta\tau}(-\mathbf{p})\).
Using the fact that
\begin{equation}
\text{det}\left[-i\omega+\mathcal{H}_{\eta\tau}(\mathbf{p})\right]=\left(-i\omega+E_{\eta\tau}(\mathbf{p})\right)\left(-i\omega-E_{-\eta\tau}(-\mathbf{p})\right)
\end{equation}
we obtain the Ginzburg-Landau free energy:
\begin{widetext}
\begin{equation}\label{F}
\mathcal{F}=-\frac{T}{2}\sum_{\mathbf{p}\eta\tau}\ln\left[2\cosh\left(\frac{\beta E_{\eta\tau}(\mathbf{p})}{2}\right)\right]-\frac{T}{2}\sum_{\mathbf{p}\eta\tau}\ln\left[2\cosh\left(\frac{\beta E_{-\eta\tau}(-\mathbf{p})}{2}\right)\right]+\mathcal{H}_{0}\left(\Delta^{2}\right)
\end{equation}
To obtain the linearized gap equation we expand \(\mathcal{F}\) to first order in \(|\Delta_{\eta\tau}|^2\), which yields
\begin{equation}\label{F2}
\mathcal{F}^{(2)}=-\sum_{\mathbf{p}\eta\tau}\frac{\tanh\left(\frac{\beta \xi_{\eta\tau}(\mathbf{p})}{2}\right)+\tanh\left(\frac{\beta \xi_{-\eta\tau}(-\mathbf{p})}{2}\right)}{4\xi_{S\eta\tau}(\mathbf{p})}|\Delta_{\eta\tau}(\mathbf{p})|^2+\mathcal{H}_{0}\left(\Delta^{2}\right)
\end{equation}
\end{widetext}
Minimizing the free energy Eq. (\ref{F2}) with respect to \(\Delta_{\eta\tau}^*(\mathbf{p})\), we obtain the gap equation (\ref{Eq:Gap}).

\section{Chiral Topological Superconductivity at Large $\alpha_R$}\label{C}

%The gapless topological crystalline superconducting phase described above exists in both singlet- and triplet- instability regimes, and hence is expected to be a generic property of superconducting monolayer NbSe$_2$ with sufficiently large in-plane magnetic fields.
As we discussed in the main text, at zero magnetic field and in the dominant triplet instability regime, at large $\alpha_Rp_F$ the leading superconducting instability is in the two-dimensional $E$ irrep of the point group $C_{3v}$. Here we show that this state spontaneously breaks time-reversal symmetry (TRS)  and compute its Chern number to show that it is a chiral topological state.

\subsection{Spontaneous Time-Reversal Symmetry Breaking}

We begin by observing that for $b=0$, and assuming equal densities of states on inner and outer Fermi surfaces, % the different $\mu$ in 
 the reduced gap equation (\ref{Eq:Gapsols}) can be solved analytically, giving: %are not coupled, so \(\Delta_{\eta,\tau}^{(\mu)}(\mathbf{p})=D_{\eta\tau}^{(\mu)}Q_{\eta,\tau}^{(\mu)}(\mathbf{p})\) are themselves solutions of the gap equation for each \(\mu=0,x,y,z\).  Explicitly, for $b=0$ they are
\begin{eqnarray}
\Delta_{\Gamma\tau}^{(0)}(\mathbf{p})&=&\tau ie^{-i\theta}D_{\Gamma}^{(0)} \\
\Delta_{\pm K\tau}^{(0)}(\mathbf{p})&=&\tau ie^{-i\theta}D_{K}^{(0)} \nonumber\\\nonumber
\Delta_{\Gamma\tau}^{(z)}(\mathbf{p})&=&\sqrt{2}ie^{-i\theta} \cos^2 3\theta\frac{\lambda p_F^3}{\delta_{\eta}(\mathbf{p})}D_{\Gamma}^{(z)}\nonumber\\
\Delta_{\pm K\tau}^{(z)}(\mathbf{p})&=&\pm ie^{-i\theta}\frac{\beta_I}{\delta_{\eta}(\mathbf{p})}D_{K}^{(z)}\nonumber\\
\Delta_{\Gamma\tau}^{(x)}(\mathbf{p})&=&\sqrt{2}ie^{-i\theta}\sin\theta\ \cos3\theta\frac{\alpha_{R}p_{F}}{\delta_{\eta}(\mathbf{p})}D_{\Gamma}^{(x)} \nonumber\\
\Delta_{\pm K\tau}^{(x)}(\mathbf{p})&=&\pm ie^{-i\theta}\sin\theta\ \frac{\alpha_{R}p_{F}}{\delta_{\eta}(\mathbf{p})}D_{K}^{(x)}\nonumber\\
\Delta_{\Gamma\tau}^{(y)}(\mathbf{p})&=&\sqrt{2}ie^{-i\theta}\cos\theta\ \cos3\theta\frac{\alpha_{R}p_{F}}{\delta_{\eta}(\mathbf{p})}D_{\Gamma}^{(y)}\nonumber\\
\Delta_{\pm K\tau}^{(y)}(\mathbf{p})&=&\pm ie^{-i\theta}\cos\theta\ \frac{\alpha_{R}p_{F}}{\delta_{\eta}(\mathbf{p})}D_{K}^{(y)}\nonumber
\end{eqnarray}
At large $\alpha_R$, there is a transition from a mixed $D^{(0)},D^{(z)}$ solution to a solution with $D^{(0)}=D^{(z)}=0$, $D^{(x,y)}\neq 0$.   This solution belongs to the 2D \(E\) irrep of \(C_{3v}\), the relevant point group in this regime.  In other words, the \((x)\) and \((y)\) solutions are degenerate, i.e. have the same \(T_c\).   %We therefore associate these two solutions with $p_{x}$-wave and $p_{y}$-wave states. 

This degeneracy opens the possibility of spontaneous  time-reversal symmetry breaking at \(b=0\).
%{\color{blue}To show that TRS is broken, we first need to express it's action on the gap functions in (\ref{GapsNoB1}-\ref{GapsNoB2}). In particular,}
In the SOC basis (\ref{Cbasis}), TRS acts as 
\begin{eqnarray}  \label{TRS}
 c_{\eta,\mathbf{p}\tau} &\stackrel{\mathcal{T}}{\rightarrow}& i\tau e^{i\theta}c_{-\eta,\mathbf{-p}\tau}  \\
 \Delta_{\eta\tau}(\mathbf{p})c^\dagger_{\eta,\mathbf{p}\tau}c^\dagger_{-\eta,-\mathbf{p}\tau} &\stackrel{\mathcal{T}}{\rightarrow} &-e^{-2i\theta}\Delta^*_{\eta\tau}(\mathbf{p})c^\dagger_{\eta,\mathbf{p}\tau}c^\dagger_{-\eta,-\mathbf{p}\tau}  \nonumber  .
 \end{eqnarray} 
 Taking \(\Delta_{\eta\tau}(\mathbf{p})=e^{i\Phi_{\eta\tau}(\mathbf{p})}|\Delta_{\eta\tau}(\mathbf{p})|\), TRS is therefore satisfied when \(e^{i\Phi_{\eta\tau}(\mathbf{p})}=\pm ie^{-i\theta}\).

In the linearized gap equation (\ref{Eq:Gap}), since the \(\mu = x\) and \(\mu = y\) channels are degenerate (i.e. they have equal critical temperatures), in principle the relative amplitudes and phases of $D_{\eta}^{(x)}$ and $D_{\eta}^{(y)}$ are not fixed (i.e. any linear combination of the two solutions is allowed). This is no longer the case if we consider the non-linear gap equations. Alternatively, in terms of the Ginzburg-Landau free energy (\ref{F}), the relative amplitudes and phases are fixed by the quartic terms in the gap functions.
%To check whether TRS is spontaneously broken, we now determine whether $D_{\eta}^{(x)}$ and
%$D_{\eta}^{(y)}$ are simultaneously non-zero, and if so, what their relative phase
%is.  To do so, we need to examine the quartic term of the Ginzburg-Landau free energy. 
Since $b=0$, $\xi_{\eta\tau}(-\mathbf{p})=\xi_{\eta\tau}(\mathbf{p})$, the free energy simplifies to
\begin{equation}
\mathcal{F}=-T\sum_{\mathbf{p}\eta\tau}\ln\left[2\cosh\left(\frac{\beta E_{\eta\tau}(\mathbf{p})}{2}\right)\right]+\mathcal{H}_{0}\left(\Delta^{2}\right)
\end{equation}
%where in the last step we assumed $b=0$, so $\xi_{\eta\tau}(-\mathbf{p})=\xi_{\eta\tau}(\mathbf{p})$.
%The contribution $\mathcal{H}_{0}\left(\left|\Delta_{\eta\tau}\right|^{2}\right)$
%comes from the decoupling of the interaction; because it is purely
%quadratic in the gaps, it is inconsequential for our analysis.
Expanding %the free energy
in powers of the gap function,
we obtain the fourth order correction (in addition to (\ref{F2})):
\begin{equation}
\mathcal{F}^{(4)}=\frac{7\zeta\left(3\right)}{64\pi^{2}T^{2}}\sum_{\eta\tau}\int N_{\eta\tau}\left|\Delta_{\eta\tau}(\mathbf{p})\right|^{4}\frac{d\theta_{\mathbf{p}}}{2\pi}
\end{equation}
where $\zeta(x)$ is the Riemann zeta function. Substituting the general
form of the gap function in the \((x)\),\((y)\) channel:

\begin{eqnarray}
\Delta_{\Gamma\tau}(\mathbf{p})&=&\sqrt{2}ie^{-i\theta}\cos3\theta\frac{\alpha_{R}p_{F}}{\delta_{\eta}(\mathbf{p})}\left(D_{\Gamma}^{(x)}\cos\theta+D_{\Gamma}^{(y)}\sin\theta\right) \nonumber \\
\Delta_{\pm K\tau}(\mathbf{p})&=&\pm ie^{-i\theta}\frac{\alpha_{R}p_{F}}{\delta_{\eta}(\mathbf{p})}\left(D_{K}^{(x)}\cos\theta+D_{K}^{(y)}\sin\theta\right) \nonumber 
\end{eqnarray}
and approximating $\frac{\alpha_{R}p_{F}}{\delta_{\eta}(\mathbf{p})}\approx1$
(which is valid as long as $\alpha_{R}p_{F}\gg\lambda p_{F}^{3}$),
we obtain:
\begin{align}
\mathcal{F}^{(4)} & =\frac{7\zeta\left(3\right)}{2048\pi^{2}T^{2}}\sum_{\eta\tau}N_{\eta\tau}\left[3\left(\left|D_{\eta}^{(x)}\right|^{2}+\left|D_{\eta}^{(y)}\right|^{2}\right)^{2}\right.\\
 & \left.-4\left|D_{\eta}^{(x)}\right|^{2}\left|D_{\eta}^{(y)}\right|^{2}\sin^{2}\phi_{xy}\right]\nonumber 
\end{align}
where $\phi_{xy}$ is the relative phase between $D_{\eta}^{(x)}$
and $D_{\eta}^{(y)}$. Minimization gives $\phi_{xy}=\pm\frac{\pi}{2}$,
which implies that the superconducting gap has the form
\begin{eqnarray} \label{Eq:TrsbGap}
\Delta_{\Gamma\tau}(\mathbf{p})&\propto&\sqrt{2}i \cos3\theta\frac{\alpha_{R}p_{F}}{\delta_{\eta}(\mathbf{p})} e^{-i ( \theta \mp  \theta)}  \nonumber \\
\Delta_{\pm K\tau}(\mathbf{p})&\propto&\pm i \frac{\alpha_{R}p_{F}}{\delta_{\eta}(\mathbf{p})} e^{-i ( \theta \mp  \theta)} 
\end{eqnarray}
which is not invariant under the time-reversal symmetry transformation (\ref{TRS}). We thus find that time reversal is spontaneously broken, in agreement with the general result in \cite{Samokhin15,Schmalian17}.
Note that while  \(\Delta_{\Gamma\tau}(\mathbf{p})\) obtained from our calculation is nodal, there is an additional symmetry allowed term $\Delta^{(3)}=e^{3i\theta}$
that belongs to the same $E$ irreducible representation.  Adding this term lifts the nodes and results in a fully gapped, time-reversal symmetry broken phase.
 
 \subsection{Chern Number and Chiral Topological Superconductivity}

To show that this TRS-breaking phase is indeed chiral, we calculate the Chern number, given by
\begin{equation} \label{ChernNo}
Ch=%\frac{1}{2\pi}\int_{BZ}F_{\eta\tau}(\mathbf{p})d^{2}p=
\frac{1}{2\pi} \sum_{\eta\tau} \int_{BZ}\mathbf{F}_{\eta\tau}(\mathbf{p})\cdot d^{2}\mathbf{p}
\end{equation}
where the  
Berry curvature vector is given by
\begin{equation}
\mathbf{F}_{\eta\tau}(\mathbf{p})=\boldsymbol{\nabla}\times\mathbf{A}_{\eta\tau}(\mathbf{p})
\end{equation}
with $\mathbf{A}_{\eta\tau}(\mathbf{p})$ the usual Berry connection associated with the occupied band only.
For a superconductor, the Berry connection is defined in terms of the normalized eigenvectors of the BdG
Hamiltonian (\ref{BdG}): $\Upsilon_{\eta\tau}(\mathbf{p})=u_{\eta\tau}(\mathbf{p})c_{\eta,\mathbf{p}\tau}+v_{\eta\tau}(\mathbf{p})c_{-\eta,-\mathbf{p}\tau}^{\dagger}$, 
via
\begin{equation}
A_{\eta\tau}(\mathbf{p})=i\langle\Upsilon_{\eta\tau}(\mathbf{p})|\nabla_{\mathbf{p}}|\Upsilon_{\eta\tau}(\mathbf{p})\rangle.
\end{equation}
In our case, the $c_{\eta,\mathbf{p}\tau}$ operators may carry a nontrivial Berry phase due to the changing orientation of the associated spin.
One should therefore consider $|\Upsilon_{\eta\tau}(\mathbf{p})\rangle$
as a four component eigenvector in a basis of Nambu-Gor'kov 4-spinors
$\Psi_{\eta\tau}^{(4)}(\mathbf{p})=(d_{\eta,\mathbf{p}\uparrow},d_{\eta,\mathbf{p}\downarrow},d_{-\eta,-\mathbf{p}\uparrow}^{\dagger},d_{-\eta,-\mathbf{p}\downarrow}^{\dagger})^{T}$.
Using the change of basis (\ref{Cbasis}), 
we  find 
\begin{equation}
\left|\Upsilon_{\eta\tau}\right\rangle =\left(\begin{array}{c}
U_{\eta\tau}^{1}(\mathbf{p})u_{\eta\tau}(\mathbf{p})\\
U_{\eta\tau}^{-1}(\mathbf{p})u_{\eta\tau}(\mathbf{p})\\
U_{-\eta\tau}^{1*}(-\mathbf{p})v_{\eta\tau}(\mathbf{p})\\
U_{-\eta\tau}^{-1*}(-\mathbf{p})v_{\eta\tau}(\mathbf{p})
\end{array}\right).
\end{equation}
 Thus  %$u_{\eta\tau}$ and $v_{\eta\tau}$ are easily found to be 
\begin{align}
u_{\eta\tau}(\mathbf{p})=\frac{\xi_{S\eta\tau}-E_{\eta\tau}(\mathbf{p})}{\sqrt{\left(\xi_{S\eta\tau}-E_{\eta\tau}(\mathbf{p})\right)^2+\left|\Delta_{\eta\tau}(\mathbf{p})\right|^{2}}} \\
v_{\eta\tau}(\mathbf{p})=\frac{\Delta_{\eta\tau}(\mathbf{p})}{\sqrt{\left(\xi_{S\eta\tau}-E_{\eta\tau}(\mathbf{p})\right)^2+\left|\Delta_{\eta\tau}(\mathbf{p})\right|^{2}}}
\end{align}
where we use the notation of Eq. (\ref{Eq:BdGspec}-\ref{Eq:Xi}). 

Below we calculate the Chern number for \(b=0\) and non-zero \(\alpha_R\) only, in which case $U_{\eta\tau}^{1}(\mathbf{p})=-i \left|U_{\eta\tau}^{1}(\mathbf{p})\right|e^{-i\theta_{\eta, \mathbf{p}}}$ where $\theta_{\eta, \mathbf{p}}$ is the angle of the momentum ${\bf p}$ measured relative to the center of the Fermi pocket $\eta$.  
Defining as before $\Delta_{\eta\tau}(\mathbf{p})=\left|\Delta_{\eta\tau}(\mathbf{p})\right|e^{i\Phi_{\eta\tau}(\mathbf{p})}$, 
we find that in this regime the Berry connection associated with the pocket $\eta$ is
\begin{equation} \label{AEq}
A_{\eta\tau}(\mathbf{p})=\left|U_{\eta\tau}^{1}(\mathbf{p})\right|^{2}\nabla\theta_{\eta \mathbf{p}}-\left|v_{\eta\tau}(\mathbf{p})\right|^{2}\left(\nabla\Phi_{\eta\tau}(\mathbf{p})+\nabla\theta_{\eta \mathbf{p}} \right)
\end{equation}
%At \(b=0\), up to a constant phase \(\phi=\theta\), the angle along the FS. From (S14), we see that for the \((0)\) and \((z)\) solutions, \(\Phi_{\eta\tau}=-\theta\) in this case, while 
For the two TRS-breaking gaps given in Eq. (\ref{Eq:TrsbGap}), we have \(\Phi_{\eta\tau}=0\)  and \(\Phi_{\eta\tau}=-2\theta_\eta\), respectively.

To obtain the Chern number we insert these expressions into (\ref{AEq}), and integrate over an annulus around each component of the Fermi surface.   
To evaluate this integral, we assume with no loss of generality that the gap function is constant in some region around the FS, and completely vanishing in regions sufficiently far from the FS, with a phase independent of the radial direction \(p\), and take \(\left|U_{\eta\tau}^{1}(\mathbf{p})\right|\) to be independent of $p$.  Finally, observe that \(v_{\eta\tau}\) changes rapidly from 0 to 1 in the vicinity of the Fermi surface.  For the pocket $\eta$, we therefore obtain:
\begin{equation}Ch_{\eta} = \frac{1}{2\pi}\int\left(F_{\eta\tau}(\mathbf{p})\right)_{p\theta}dp\ d\theta=\frac{1}{2\pi}\int\partial_p\left(A_{\eta\tau}(\mathbf{p})\right)_{\theta} dp\ d\theta\end{equation}
\begin{equation}=\frac{1}{2\pi}\left[\int \left(A_{\eta\tau}(\mathbf{p})\right)_{\theta}  d\theta \right]_{p=0}^{p=\infty}=-\frac{1}{2\pi}\left[\Phi_{\eta\tau}(\mathbf{p})+\theta_{\eta, \mathbf{p}} \right]_0^{2\pi}\end{equation}
%{\fb Changed the last $\phi$ to a $\theta$ -- Daniel please check.  ($\phi$ had not been defined; I think this was a typo?}  
where the integrals over $\theta$ and $p$ are understood to be over the tangential and normal directions in a disk including the Fermi surface of the $\eta$ pocket, respectively.  This gives a net Chern number of \(\pm6\), with a total contribution of \(\pm4\) from the $\pm K$ pockets, and of \(\pm2\) from the $\Gamma$ pocket. 

We emphasize that this result is independent of the choice of  \(\Delta_{\eta\tau}\) away from the Fermi surface.   This is because only the region proximate to the Fermi surface contributes to the integral in Eq. (\ref{ChernNo}).   %the result does not depend on how the band structure of the BdG Hamiltonian is completed far from the Fermi surface. 
 To see this, notice that any two choices of the superconducting gap away from the Fermi surface must yield the same result, as a topological phase transition requires band touching that can only occur when \(\xi_{S\eta\tau}\) and \(\Delta_{\eta\tau}\) are simultaneously zero in Eq. (\ref{Eq:BdGspec}) (as can be directly verified by constructing a simple homotopy between Hamiltonians with any two such choices).

\section{Tight-binding model on a cylinder} \label{TightBinding}

To study the edge modes in more detail and produce the plot in Fig. \ref{CylFig} we used a tight binding model defined on the triangular lattice. % with sites we label with indices \(i\) and \(j\), which stand for \(\mathbf{R}_i=n_i \mathbf{a}_1+m_i \mathbf{a}_2\), where \(\mathbf{a}_1=(a,0)\) and  \(\mathbf{a}_2=\frac{a}{2}(1,\sqrt{3})\) are the lattice basis vectors. We take \(a=1\) for convenience below, and also define \(\mathbf{a}_3=\mathbf{a}_2-\mathbf{a}_1=\frac{a}{2}(-1,\sqrt{3})\). Each site then has six nearest neighbors at \(\pm\mathbf{a}_1\), \(\pm\mathbf{a}_2\) and \(\pm\mathbf{a}_3\). 
The Hamiltonian has the general form
\be
H = H_0 + H_Z + H_{SC} %=\sum_{ij\alpha\beta} \left(\mathcal{H}_0\right)^{ij}_{\alpha\beta}d^\dagger_{i\alpha}d_{j\beta}
\ee
%{\fb Changed notation here to keep $H_0$ consistent with main text.  Commented out expression involving creation operators as it was not correct.}
The first term describes the normal state band structure in the presence of SOC; the second-term is the Zeeman coupling due to in-plane magnetic field; and the last term represents the superconducting pairing gap.  For simplicity we use a tight-binding model that only includes the \(\eta=\Gamma\) pocket Fermi surface, since the $\pm K$ pockets are unimportant for the crystalline nodal topological superconductor. 

We describe our model in terms of the creation operators $d^\dag_{i, \alpha}$, where $\alpha =\uparrow, \downarrow$ is a spin index, and $i$ is a site index.  We have
\ba
H_0&=&\sum_{i\alpha}\mu\ d^\dagger_{i\alpha}d_{i\alpha}+\sum_{\langle ij\rangle\alpha}t\ d^\dagger_{i\alpha}d_{j\alpha} \n
 & & + \sum_{\langle ij\rangle\alpha\beta}\left[4i\lambda\nu_{ij}\sigma^{z}_{\alpha\beta}+\frac{i\alpha_R}{3} \hat{{\bf z}} \cdot \left( \boldsymbol{\sigma} \times {\bf a}_{ij} \right) _{\alpha\beta}\right]d^\dagger_{i\alpha}d_{j\beta} \nonumber \\
H_{Z}&=&\sum_{i\alpha\beta}\left(\mathbf{b}\cdot\boldsymbol{\sigma}\right)_{\alpha\beta} d^\dagger_{i\alpha}d_{i\beta}\\
%The SOC Hamiltonian has been described in \cite{KaneMele} (with the slight proviso that they had a hexagonal lattice), which is
%H_{SOC}&=
H_{SC}&=&\frac{1}{2}\sum_{ij\alpha\beta}\left[\Delta\right]^{ij}_{\alpha\beta}d^\dagger_{i\alpha}d^\dagger_{j\beta}+\text{h.c.} \nonumber
\ea
 where $\mathbf{a}_{ij} \in \{ \pm \mathbf{a}_1, \pm \mathbf{a}_2,\pm \mathbf{a}_3 \} $ is the vector from site \(i\) to site \(j\), and \(\nu_{ij}=1\ ( -1)\) if the vector is \(\mathbf{a}_1\), \(-\mathbf{a}_2\), \(\mathbf{a}_3\) ( \(-\mathbf{a}_1\), \(\mathbf{a}_2\), \(-\mathbf{a}_3\)) .   For our triangular lattice, \(\mathbf{a}_1=(a,0)\) and  \(\mathbf{a}_2=\frac{a}{2}(1,\sqrt{3})\),  \(\mathbf{a}_3=\mathbf{a}_2-\mathbf{a}_1=\frac{a}{2}(-1,\sqrt{3})\).
 We consider the singlet-instability regime, in the crystalline nodal topological phase where $b \gg \alpha_R$.  In this region the self-consistent solutions of the gap equation obtained in a \(\mathbf{k\cdot p}\) model are well-approximated by
\be \label{GapFunc}
\Delta^{ij}=\Delta_t \nu_{ij} \left(\sigma^x\cos\vartheta+\sigma^y\sin\vartheta \right)i\sigma^y+\Delta_s i\sigma^y
\ee
where  \(\vartheta\) is the direction of the magnetic field, assuming \(\Delta_s \ll \Delta_t\) (higher lattice harmonics are in general needed to match the \(\mathbf{k\cdot p}\) model exactly).
The numerical coefficients are chosen  to match the \(\mathbf{k\cdot p}\) Hamiltonian (including the value of $p_F$). %at the chosen values of $b$ and $\alpha_R$. 
%{\fb Clarified the previous sentence -- Daniel please check.}
%The specific values used in the Figure are listed in Section \ref{parameters}.

%Finally, taking
%\begin{equation}H_{SB}=H_0+H_{SOC}+H_Z=\sum_{ij\alpha\beta} \left(\mathcal{H}_0\right)^{ij}_{\alpha\beta}d^\dagger_{i\eta\alpha}d_{j\eta\beta}\end{equation}
%the tight binding BdG Hamiltonian on the real lattice is given by
%\begin{equation}H_{BdG}=\frac{1}{2}\sum_{ij}\Psi_{i}^\dagger \mathcal{H}\Psi_{j}\end{equation}
%where \(\Psi_i=\left(d_{i\uparrow},d_{i\downarrow},d^\dagger_{i\uparrow},d^\dagger_{i\downarrow}\right)\) and
%\begin{equation}\mathcal{H}=\left(\begin{array}{cc}
%\mathcal{H}_0^{ij} & \Delta^{ij}\\
%-\Delta^{ij*} & -\left(\mathcal{H}_0^T\right)^{ji} 
%\end{array}\right)\end{equation}
%(the transpose is in spin indices only).

Our cylinder is created by taking periodic boundary conditions in the vertical \(y\) direction, and open zig-zag boundary conditions along the \(x\) direction.  To produce the plot, we Fourier transform in the $y$ direction: 
\begin{equation}d_{\mathbf{R_i}\alpha}=\frac{1}{\sqrt{N}}\sum_{p_y}d_{R_{ix}p_y\alpha}e^{-ip_yR_{iy}}\equiv\frac{1}{\sqrt{N}}\sum_{p_y}d_{i p_y\alpha}e^{-ip_yR_{iy}}\end{equation}
where \(\mathbf{R}_i=(R_{ix},R_{iy})\). Note that \(i\) labels the \(x\) coordinates of the sites which go in increments of \(a/2\), while the period along the \(y\) axis is actually doubled since identical sites are now separated by \(2\mathbf{a}_2\), resulting in the folding of the 1D Brillouin zone (which has a period of \(\frac{2\pi}{\sqrt{3}a}\)).

The resulting BdG Hamiltonian on the cylinder can be expressed
\begin{equation}H_{BdG}=\frac{1}{2}\sum_{ij,p_y}\Psi^\dagger_{i,p_y} \mathcal{H}^{ij}(p_y)\Psi_{j,-p_y}\end{equation}
where \(\Psi_{i,p_y}=\left(d_{i,p_y\uparrow},d_{i,p_y\downarrow},d^\dagger_{i,-p_y\uparrow},d^\dagger_{i,-p_y\downarrow}\right)\) and
\begin{equation}\label{BdGcyl}\mathcal{H}^{ij}(p_y)=\left(\begin{array}{cc}
\mathcal{H}_{kin}^{ij}(p_y) & \Delta^{ij}(p_y)\\
-\left(\Delta^{ij}(-p_y)\right)^* & -\left(\mathcal{H}_{kin}^T(-p_y)\right)^{ji}
\end{array}\right)\end{equation}
%We refer to the spectrum of \(\mathcal{H}^{ij}(p_y)\) as the edge spectrum.
where we have defined
\be\label{Hkin}
H_0 + H_Z =\sum_{ij\alpha\beta} \left(\mathcal{H}_{kin} (p_y) \right)^{ij}_{\alpha\beta}d^\dagger_{i p_y \alpha}d_{j p_y \beta}
\ee
In Fig. \ref{CylFig} we plot the spectrum of (\ref{BdGcyl}) with the number of sites along the non-periodic \(x\) direction \(N=300\) (which corresponds to 150 unit cells due to period doubling). We also set \(t=1\), \(\mu=0\), \(\beta_I=1\), \(\lambda=0.2\), \(b=1\), \(\alpha_R=0.1\), \(\Delta_t=1\) and \(\Delta_s=0.1\). The magnetic field was again aligned along one of the \(\Gamma\)-\(K\) directions, \(\vartheta=0\).

\end{document}